\documentclass[%
 aip,
 amsmath,amssymb,
 reprint,%
]{revtex4-1}

\usepackage{graphicx}
\usepackage{dcolumn}
\usepackage{bm}
\usepackage{url}

\usepackage[utf8]{inputenc}
\usepackage[T1]{fontenc}
\usepackage{mathptmx}
\usepackage{etoolbox}
\usepackage{bbm}
\usepackage[usenames,dvipsnames]{xcolor}   

\graphicspath{{"./figures-NLDofRC_3/"}}

\makeatletter
\def\@email#1#2{%
 \endgroup
 \patchcmd{\titleblock@produce}
  {\frontmatter@RRAPformat}
  {\frontmatter@RRAPformat{\produce@RRAP{*#1\href{mailto:#2}{#2}}}\frontmatter@RRAPformat}
  {}{}
}%
\makeatother
\begin{document}

\preprint{AIP/123-QED}

\title{Statistics for Differential Topological Properties between Data Sets with an Application to Reservoir Computers} 
\author{Louis Pecora}
 \affiliation{Institute for Research in Electronics and Applied Physics, U. Maryland}
 \email{louboog.twogrand@gmail.com}
\author{Thomas Carroll}
\affiliation{Retired}.

\date{\today}

\begin{abstract}
Abstract: Recording (multi-dimensional) time series from dynamical systems in experiments or numerical studies on coupled systems is common. In many cases we want to find relationships between systems or subsystems, but multi-dimensional data prevents simple graphical tests. We here develop statistics that test the data for basic, fundamental relationships, like the existence of continuous functions between the subsystems.  These are common fundamental questions that should be answered before diving into function fitting and other data handling schemes, since lacking continuity and, possibly, smoothness in data relations guarantees that other more specific data fitting attempts will fail or be meaningless. We develop useful testing statistics by creating data driven versions of basic topological concepts including continuity, differentiability, point set distance comparisons, diffeomorphisms, and embeddings.  As an application we show how such statistics can aid in the analysis of the dynamics of reservoir computer systems, many of which rely on there being specific relations (like embeddings) between the drive and the reservoir computer.
\end{abstract}

\maketitle

\begin{quotation}
Lead paragraph: It is common for researchers to record long, multiple time series from experiments or calculations. But sometimes there are no good models for the systems or no applicable mathematical theorems that can tell us when there are basic relationships between subsets of the time series data such as continuity, differentiability, embeddings, etc. The data is often higher dimensional and simple plotting will not guide us. At that point fitting the data to polynomials, Fourier series, etc. becomes uncertain. Even at the simplest level, having data that shows there is a function between the data subsets is useful and a negative answer means that more particular data fitting or analysis will be suspect and probably fail. We show here statistics that test time series subsets for basic mathematical properties and relations between them that not only indicate when more specific analyses are safe to do, but whether the systems are operating correctly.  We apply these statistics to examples from reservoir computing where an important property of reservoir computers is that the reservoir system establishes an embedding of the drive system in order to make any other calculations with the reservoir computer successful.
\end{quotation}

\section{ Introduction\protect}

With the advent of computers with large memories and fast operations along with sophisticated data gathering equipment in the laboratory or field we long-ago entered the age of "big data." Whether the data is gathered from measurements or numerical calculations, a lot of effort in all fields of science and applied mathematics has been put into analyzing the data sets.  One of the somewhat overlooked questions regarding data sets is, given two data sets what can we say about the possible mathematical relationships between them? 

Here we are considering very basic relationships that form the basis of analysis and calculus on multidimensional data sets.  If we are given two multidimensional data sets, which are matched point for point, what can we say about possible functions that map one to the other?  What properties can we say such functions have? Continuity? Invertibility? Differentiablity? 

Here we assume the functions are acting on a sampling of points from each data space, which are in some way dense in their spaces. In other words, they are on some manifold in their space although they do not necessarily fill out the whole manifold, i.e. they are not dense in the strict mathematical meaning of the word. Or they have associated with them a measure (as strange attractors have). 

This also brings up the necessity of examining whether two sets of points are from the same construct (e.g attractor). Although this might seem easy to see by just plotting the data, we will see that when one is dealing with high-dimensional data sets it is not trivial and requires a good statistic based on definitions of distances between sets of points in a metric space.

All the above qualities of possible functions between data sets are defined in simple differential topological definitions for relationships between sets of points in metric spaces \cite{Chillingworth-DiffTop}. 

What does all this have to do with reservoir computing systems? The main link there is provided by Takens' embedding theorem \cite{Takens-EmbTheo} and work by Sauer, Yorke, and Casdagli \cite{SauerYCprev} and Hunt, Sauer, and Yorke \cite{HuntSYprev2}.  We will show a simplified explanation of how attractors on manifolds in dynamical systems driving a reservoir computer can (under the right circumstances) provide an embedding of the drive system in the reservoir system.  The tests for this possibility using only data will involve the statistics we mentioned above. To get to this point we start by introducing reservoir systems and clarifying which principles are important for understanding their dynamics as well as which explanations of their behavior are misleading.

\section{ Reservoir Computing\protect}

Reservoir computing was introduced in the early 2000's in a few papers \cite{Jaeger:2001aa, Mass:2002, Jaeger:2007, Jaeger:2004}, which influenced the machine learning field more than the nonlinear dynamics field, at least at first. To put it simply, it was initially shown that a reservoir computer (RC) could be driven by an input of one component of a dynamical system (the drive) and it could be trained to reproduce the other (unseen by the reservoir) drive components. And, once trained it could do this accurately for signals from other initial conditions of the same drive system. 

 There were some early explorations by dynamics researchers \cite{Parlitz:2005}, but mostly reservoir computing was ignored until some years later (for example see refs.\cite{Appeltant:2011,PathHunt,LuHunt,LuPath}). However, after 2010 or so the topic of "reservoir computing" grew in size rapidly as people in the field of nonlinear dynamics realized that the success of these devices had to be related to the dynamics of the devices since a reservoir computer (RC) was simply a network of coupled nonlinear dynamical nodes.
 
In a sense we are looking at the observer problem, using a reservoir computer to reproduce or predict a signal. The requirements for classifying different signals using a reservoir computer will be different. 

Even with the emergence of RCs in the nonlinear dynamics world, some of the explanations of how the RCs worked and what properties were important to their functioning were influenced by loose concepts or simply historical precedence (see, for example: \cite{AntSch,LukosM}). These include ideas like the nodes should be sigmoid functions like many neural networks, the operation of the reservoir computer should be as near to the unstable point in the system's dynamical parameters as possible, and the recurrent, coupling network linking the dynamics of all nodes should be randomly chosen and sparse \cite{Nakajima-Fischer}. The claim about operation of the reservoir computer near the edge of chaos has been shown to not be universal \cite{Carroll-Edge_ofChs} and we will enhance the dynamics in that case here beyond the references.  The latter claim about randomness makes some sense since special networks like those with symmetries can cause lower dimensional dynamics \cite{Pecora:2014, Nicosia2013, Sorrentino2016} and negatively effect performance of reservoir computers \cite{Carroll:2019}. 

Recently, in a forward to a book on reservoir computing \cite{Nakajima-Fischer}, H. Jäger has clarified some of the interesting history of Echo State and Liquid State systems (now subsumed under the name Reservoir Computers). In this were also noted some of the misconceptions of these systems and their (sometimes puzzling) origin, like operating at the "edge of chaos."

Herein we focus on identifying the correct dynamical behavior that we are seeing in RCs that are successfully performing their machine learning tasks. Many of these are related to dynamical behavior or properties of operating RCs, but are very hard to prove or establish mathematically. For example, in the case of a drive signal from an input that is a dynamical system, one of these properties is that a properly performing RC is usually generating an embedding of the attractor of the drive signal. This property can be proven for some cases rigorously \cite{GrigoryevaOrtega2021}, but is very difficult to prove for many RC systems. And, of course, impossible to prove in many cases of analog systems where the equations of motion and the node properties are not well-known.  

We address these problems by developing statistics like those mentioned in the introduction that capture dynamical, functional, and geometric properties of the RCs. All of our statistics involve some assumption that the RC is operating dynamically, so it turns out even simple statistics have extensions that will hold under dynamical behavior. 

Throughout this paper we will also mention other questions that need to be studied beyond what we present here in detail. We will start by a detailed description of RCs. For a more dynamical view of RC see also the nice review of RC by Parlitz \cite{ParlitzDelays}

\section{\label{sec:level1} Basic structure of Reservoir Computer and Test Systems\protect}

\begin{figure}
\includegraphics[ width=\columnwidth,height=0.21\textheight]{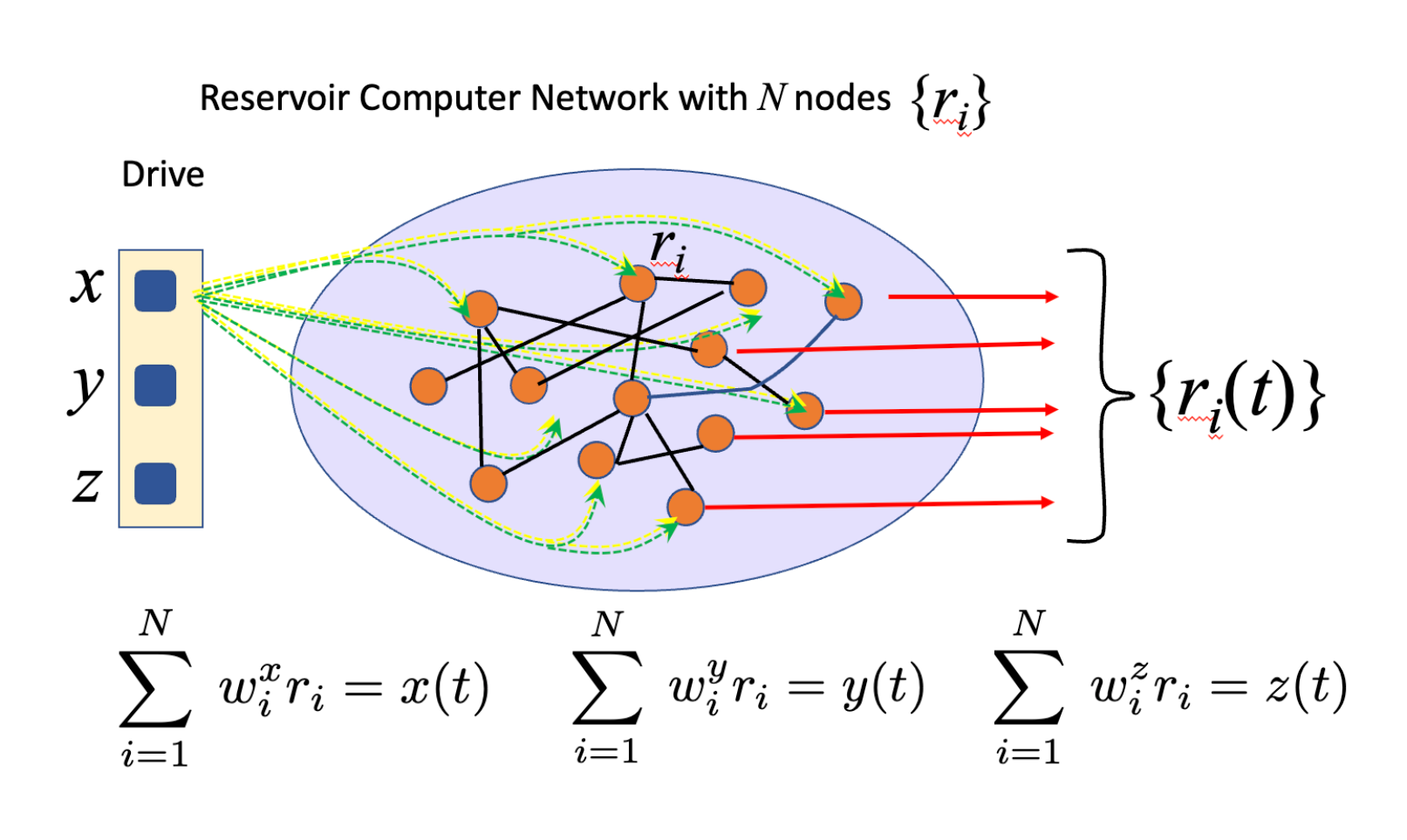}
\caption{\label{fig:RC-Schematic} RC Schematic: A RC driven by the $x$ component of a 3D system with the $y$ and $z$ components reconstructed from the RC node dynamical variables $r_i(t)$ using their respective weights $w_i^y$ and $w_i^z$ calculated from a training run using the original drive, but with different time series. Here we assume that the RC dynamics of each of the $N$ nodes are 1-dimensional for simplicity. This is generalized in the text.}
\end{figure}

The general layout of a RC being driven by a component of a dynamical system with the output reconstructing other dynamical variables is shown in Fig. \ref{fig:RC-Schematic}. We will refer to the input to the RC as the drive. For this paper the drive is an autonomous dynamical system. The input can also be a function of a component of the drive, but it must be an invertible function to eventually be part of an embedding of the drive attractor. 

The RC is a coupled network which provides signals from some other nodes to each node. This is often referred to as a {\em recurrent} network. Its structure in this paper will be a random network. This is described more below. Signals from the drive are put into each node or subset of nodes.  Herein, the input will be to each node, but the amplitudes of the inputs will vary between nodes in the RC, usually with random input amplitudes selected from a uniform distribution on $[-1,1]$. 

We assume we are using simultaneously recorded time series from the drive and the RC. We start with the assumption that all dynamical variables are recorded in the drive and the RC.  

After a transient period of input the nodes outputs are fed to an output analyzer/computer that at each time step maps the output onto values of many or all of the components of the original drive at that time step.  This output can be a combination of various functional forms, but here we will use only a linear mapping, which is generally what most research studies of RCs do, including the original papers on the topic. The training of the RC involves only learning the parameters ($w_i^x$, etc. in FIG. \ref{fig:RC-Schematic}) of the linear map from the RC outputs to the drive variables. The training does not involve changing anything in the RC, making it fast and effective. In the end, if successful, an RC driven by one of the drive's variables can have its output mapped to all the internal dynamical variables of the drive, i.e. it is reconstructing the unseen drive components.

It is also possible to have the RC predict the drive dynamical variables one step ahead in time and then feed back those values into the RC to generate the drive system's variables another step ahead in time, etc.  This way the RC can be trained to predict the drive variables and serve as an autonomous dynamical system itself that emulates the original dynamics of the drive system on it's own when the original drive system is removed.

This process does not necessarily work for any input from the drive.  Which ones are successful varies from drive to drive. The statistics we develop here can be useful in determining which combinations of drive signals work best. The outputs of the RC are signals from the nodes of the RC that are from the dynamical variables of the nodes.  We extract all nodes' signals, but it's still somewhat an open question of which nodes of a subset might work just as well. 

In general, it is easy to see that the number of hyper-parameters involved in RC is huge. An obvious sampling of them include the network adjacency or coupling matrix (structure, orientations, and weights), types of and parameters of the RC nodes, input weights and selected inputs to the RC nodes from the drive, and coupling types and parameters between the RC nodes. Several of the aforementioned have multiple parameters and type choices within them. Some optimization algorithms have been applied to these, but full optimization remains an onerous task. 

A more judicious approach might be to find those parameters that are most important to correct operation of the RC for some subset of inputs. For example, it will probably be necessary to have the RC and the drive operate in the same frequency band. Other parameters may be only slightly influential in getting the RC to operate in an advantageous way. We do not address all these questions here, but our approaches will be helpful in answering the questions we just raised.


\section{\label{sec:level2} Dynamics and Embeddings in a drive-\\reservoir system\protect}

For our explanation of why the RC can reproduce an embedding of the drive system we will first concentrate on dynamical systems that are maps or iterative dynamics. These have simpler mathematical expressions than flows. After that we will move to ordinary differential equations (ODEs) and express results as flows. For a more rigorous introduction to the mathematical formulation of nonautonomous systems in RC, see Ceni et. al \cite{CeniAsh}.

We write the autonomous drive system as,
\begin{equation}
  \label{eq:drvDyns}
  {\bf{x}}(n+1)= {\bf{f}}({\bf{x}}(n)),
\end{equation}
where ${\bf{x}}\in {\mathbbm{R}}^K$ and ${\bf{f}}: {\mathbbm{R}}^K \rightarrow {\mathbbm{R}}^K$.

We select one input variable to the RC ($x_i$, say, for some particular value of $i \in \{1,...,K\}$). In general, we can also use an invertible function $u(x_i)$ as the drive input to the RC.  For now, though, we will keep things simple and use $u(x_i)$= identity function in our calculations, but keep the $u$ in our mathematical expressions for the most part to show the generality. Note, that we can also use a subset of multiple drive components, too, if we are careful of how we define a function of them.  This also requires careful definition of the $u$ function and we will avoid this for now by using only one component of the drive. 

With $u(n)=x_i(n)$ we assume the RC has $N$ nodes each node having a dynamical dimension of $M$. We then write the dynamics of the nonautonomous RC as,
\begin{equation}
  \label{eq:RCDyns1}
  {\bf{r}}(n+1)= {\bf{F}}(u(n),{\bf{r}}(n)),
\end{equation}
where ${\bf r} \in {\mathbbm R}^{NM}$. Using function composition iteratively in Eq.(\ref {eq:RCDyns1}) and iterating on ${\bf r}(n)$ we can write Eq.(\ref {eq:RCDyns1}) in an insightful way \cite{Jungling_privateComm_1}. That is, we write, 
\begin{equation}
  \label{eq:RCDyns2}
  {\bf{r}}(n+1) = {\bf{F}}[u(n),{\bf{F}}[u(n-1),{\bf{F}}[(n-2),{\bf{F}}[u(n-3),...]
\end{equation}
From the above Eq.(\ref {eq:RCDyns2}) we see that the "next" time step ($n+1$) depends iteratively on the previous $n$ steps, assuming $n=0$ is the initial time and ${\bf{r}}(0)$ and ${\bf{x}}(0)$ the initial conditions. Eq.(\ref {eq:RCDyns2}) is reminiscent of the form of Taken's theorem \cite{Takens-EmbTheo} with respect to a dependence of the dynamics reconstruction on multiple time shifted terms. 

Briefly, Takens\cite{Takens-EmbTheo} proved that for many dynamical systems one can reconstruct an entire attractor from a single measured time series quantity (even only one variable). This is done by using the measured variable (say, $x(t_i)$ for many time points $\{t_i|i=1,2,...\}$) and constructing the $n$-dimensional vectors ${\bf v}(t_i)= (x(t_i),x(t_{i-1}),x(t_{i-2}),...,x(t_{i-(n-1)}))$. The points ${\bf v}(t_i)$  are then points on the attractor and this reconstruction is diffeomorphic to the actual attractor for non-chaotic systems. Since most chaotic attractors and some non-chaotic ones\cite{Grebogi-NonCht} are fractal objects it is necessary to extend Takens proof using the concept of prevalence\cite{HuntSYprev2,SauerYCprev}, which allows us to treat the Takens time-delay reconstruction as part of a manifold itself.  

We can show that with a few more general assumptions Taken's theory reconstruction indeed does apply here to RCs. In the case of continuous time evolution of the RC and drive we can get the same type of result as for Eq.({\ref {eq:RCDyns2}}), although it is a little more involved. Let $\psi_t$ be the flow of the drive system and $\Psi_t$ be the flow of the RC. The dynamics of the system is in the form of a skew product flow \cite{KloedenYangBOOK}.  That is the drive $\psi_t$ is autonomous, but the RC $\Psi_t$ is nonautonomous. The important property we will need is the cocycle property \cite{KloedenYangBOOK} of the flow on a nonautonomous system (the RC): $\forall \: s,t \in \mathbbm{R}^+ \text{, } \bf{x}$  in the drive space, and $\bf {r}$  in the RC space, the following describes the flow:

\begin{equation}
\begin{aligned}
  \label{eq:RCflow}
  {\bf{r}}(t+s) &= \Psi_{t+s}[{\bf {x}}, {\bf {r}}] \\
  &= \Psi_{t}[\psi_{s}[{\bf {x}}],\Psi_{s}[{\bf {x}},{\bf {r}}]] 
\end{aligned} 
\end{equation}

Then, iterating the above drive-RC embedding equations analogously to Eq.(\ref{eq:RCDyns2}),  Eq.(\ref {eq:RCflow}) can be written as,
\begin{equation}
\begin{aligned}
  \label{eq:RCflow3}
  {\bf{r}}(t) &= \Psi_{t}[{\bf {x}}_0, {\bf {r}}_0] \\
  &= \Psi_{\tau}[\psi_{t-\tau}[{\bf {x}}],\Psi_{t-\tau}[{\bf {x}}_0, {\bf {r}}_0]]\\
  &= \Psi_{\tau}[\psi_{t-\tau}[{\bf {x}}],\Psi_{\tau}[\psi_{t-2\tau}[{\bf {x}}],\Psi_{t-2\tau}[{\bf {x}}_0, {\bf {r}}_0]]] \\
  &= \Psi_{\tau}[\psi_{t-\tau}[{\bf {x}}],\Psi_{\tau}[\psi_{t-2\tau}[{\bf {x}}],
\Psi_{3\tau}[\psi_{t-3\tau},[{\bf {x}}], \Psi_{t-2\tau}{\bf {r}}_0]]]\\
\text{etc.}
\end{aligned}
\end{equation}
And if we abuse the notation a little, we see how the time delays of the drive system enter the dynamics of the RC:
\begin{equation}
\begin{aligned}
  \label{eq:RCflow4}
  {\bf{r}}(t) = 
\Psi_{\tau}[{\bf{x}}(t-\tau),\Psi_{\tau}[{\bf{x}}({t-2\tau}),
\Psi_{\tau}[{\bf{x}}(t-3\tau), \\ \Psi_{\tau},
...[{\bf {x}}_0,{\bf {r}}_0]]],
\end{aligned}
\end{equation}
where Eq.(\ref{eq:RCflow4}) and Eq.(\ref{eq:RCDyns1}) are similar to that presented by Takens\cite{Takens-EmbTheo} for attractor reconstruction.

First, we are assuming that the dimension of the RC dynamics is large enough to support an embedding of the drive and that the drive component can itself accomplish a Takens embedding (any arbitrary component of a dynamical system cannot not necessarily successfully accomplish a Takens embedding). Second, to make things explicit we write for the map case ${\bf{r}}(n)$ as the last term in a sequence of functions:  ${\bf{g}}_k({\bf r}_0)={\bf F}[u(k),{\bf F}[u(k-1),...]$ for $k\in \{0,1,..,n\}$ with a similar equation for the flow case. We have two requirements: (1) we want the sequence  to converge as $n\rightarrow \infty$ and (2) we want the sequence to become independent of ${\bf r}_0$, i.e. we want asymptotic convergence to the RC dynamics. 

To get convergence we can require the sequence 
$\{ {\bf g}_k({\bf r}_0) \}$ to obey the Cauchy condition: $\forall \: \epsilon > 0$ $\exists \:\:n \in {\mathbb N}^+$ such that $| {\bf g}_k({\bf r}_0) - {\bf g}_l({\bf r}_0) | < \epsilon \; \forall \; k,l \ge n$. This guarantees that the sequence converges and the convergence carries all the properties of the individual ${\bf{g}}_k({\bf r}_0)$ functions, i.e. continuity, differentiability, etc. We can write similar convergence conditions for the case of flows.

To get independence of the trajectory of the RC from the initial conditions $\{ {\bf r}_0 \}$, as above, we require the sequences ${\bf g}_k({\bf r}_0)$ and ${\bf g}_l({\bf r}'_0)$ to converge to the same attractor for all initial conditions, ${\bf r}_0$ and ${\bf r}'_0$ in the same basin of attraction.  

The result of these requirements is that the RC trajectory does not depend on all past trajectory points, but approximately only on a finite number of recent ones. This leads to the situation where we can get a good approximation of the image of the drive attractor in the RC trajectory by driving with a finite number of drive time series points that are sequential time delays of the drive. This just as in a Takens reconstruction of an attractor. That is, the RC trajectory is on a manifold that contains an embedding of the drive attractor.

The above explanation does not amount to a strict mathematical proof of the connection between the RC trajectory and some diffeomorphism of the drive attractor. We simply want to show that such a relation is reasonable and even likely. For more rigorous proofs of the RC-drive attractor relationship see the papers \cite{Hart-EmbdgEchoStates,GrigoryevaOrtega2021}.  Proofs for the theorems in these papers  are at times limited to drive and RC systems with special properties (e.g. the necessity of contracting maps as a property of the dynamics). We emphasize that the above statements are not criticisms of the referenced papers.  The proofs are involved and highly non-trivial.  But so far the research on RCs suggests that the actual requirements for an embedding are less restrictive.

We also note that our requirements for the properties of the RC attractor trajectory leading to a finite number of terms in the essential dynamical history of the system is in the same vein as the conclusions in the seminal paper by Chua and Boyd \cite{Boyd-Chua-Volt}. In that paper they showed that a Volterra series representing the dynamics of a solution to a differential rate equation can be truncated at a finite number of terms if the system has fading memory.  That is, it forgets the past as one goes back in time from any point on the trajectory. We reach similar conclusions from a more general starting point.

We add that the fading memory property is often not quantified as much as it should be.  Like the application of Takens theorem, the number of past points that contribute to the reconstruction of the attractor matters. Too few points into the past is like an {\em underembedding} and has a similar effect as using too small a dimension for the embedding space.  The attractor in this case is not unfolded enough or, to put it another way, parts of the trajectory which should be in separate regions of the reconstruction space are projected onto other parts because the dimension is not large enough to separate the intersecting regions. Too many points in the time-sequence reconstruction results in an {\em overembedding} where the folding of the attractor as in the case of chaos is placed into the reconstruction making for a highly distorted attractor \cite{Kantz_Obrich-1997}


In the next few sections we take the view that many drive-RC systems can generate time series data that can be used to test for the likelihood that the system has certain mathematical properties related to embeddings. This requires the development of statistics that we mentioned in the introduction.



\section{\label{sec:level3} Continuity Statistic\protect}

We want a set of statistics that provides evidence that we do or do not have a diffeomorphism 
between the drive and the RC. We first assume that we have full sets of time series for both 
the drive and the RC, which are sampled simultaneously at a rate that is fast enough to capture spatially local and temporally close system behaviors. Each vector of drive points at some time $t$ has a simultaneous set of RC points in a vector sampled at the same time $t$. We will refer to these as time-twin points or just \emph {time-twins}. This allows us to use the time-twins to represent a point to point mapping between the drive and the RC. We will also mention some alternatives if all time series are not available.

In this section on continuity and in the next section on differentiability, for simplicity of explanation we will treat the process of finding spatially nearby points as one of a simple search for the closest data points. This process is actually more subtle for dynamical time series. We will address the proper way to find nearby, but time-uncorrelated points.

Searching for closest points uses a metric on both spaces which we assume to be the usual square-root of sum of squares of component differences.  This automatically translates into a set of open sets on each space as defined in the usual way by the metric and, hence, to a metric-induced topology on both drive and RC data spaces. As a result we use the same metric to define the properties of continuity, differentiability, etc. without resorting to a topological definition in terms of function inverses, open sets, etc.  The latter would add a more abstract approach, but it would be unnecessarily so.

We first define a diffeomorphism:

{\bf {Definition}}: A \textit {diffeomorphism} is a bijective map between two manifolds ${\mathcal M}$ and ${\mathcal N}$, which is continuous and differentiable and which has an inverse that is continuous and differentiable.

Note that this type of mapping preserves a lot of geometric properties between the manifolds.  Many properties of ${\mathcal M}$ are also true of ${\mathcal N}$ and vice versa. Especially important is the fact that many dynamical properties that depend on continuity and differentiability are the same for both manifolds. These include the same Lyapunov exponents and fractal dimensions of the attractors on the manifold which are also connected by the manifolds' diffeomorphism. 

Note, that many attractors are not themselves manifolds, but are contained in manifolds. As mentioned in the previous section, the prevalence extension to Takens theorem allows us to estimate local properties like continuity and differentiability, for which the latter needs a tangent space on the manifold.

The first fundamental property we consider is continuity. We would like to find a numerical/statistical test that could give us some idea if there is a continuous mapping between two simultaneously sampled attractor data sets. We first developed a continuity statistic in previous papers \cite{Pecora-ContDiff:1995, Pecora-attrRecon:2007} and we introduce a more sophisticated version here using a null hypothesis to which we will give a geometric meaning.

We start with the usual calculus definition of continuity at a particular ({\em fiducial}) point ${\bf x}_0$ in ${\mathcal M}$ for a function $\bf \phi$ mapping a subset of points in a metric space to a subset in another metric space, say $\bf \phi:{\mathcal M} \rightarrow{\mathcal N}$:
\begin{equation}
\begin{aligned}
  \label{eq:ContDef}
  \forall \epsilon > 0  &\;\; \exists \delta > 0: \text {for }{\bf x}\in {\mathcal M}\\ 
  &\text {whenever} \;|{\bf x}-{\bf x }_0|< \delta \; \Rightarrow \;|\phi ({\bf x})-\phi ({\bf x}_0)|< \epsilon \text{ ,}
\end{aligned}
\end{equation}
where we use absolute value bars to represent the metric distances. The drive will be $\mathcal M$ and the RC will be $\mathcal N$. $\mathcal M$ and $\mathcal N$ will change places when we also consider the inverse mapping, which must also be continuous for an embedding.

Of course, we have a finite amount of time series data so we cannot take the limit as $\epsilon \rightarrow 0$. At some $\delta$ or $\epsilon$ size we will run out of points.  The real question is, how many points are enough to test for continuity as in the definition above at the scale of  $\epsilon$?

We solve this problem at each fiducial point ${\bf x}_0$ by proposing a null hypothesis that assumes some probability of a $\delta$-point getting mapped into the $\epsilon$ set  as follows.  We pick a $\delta >0$ and look at where the points in $\cal M$ inside the $\delta$ set get mapped to in $\cal N$. We then see what size $\epsilon$ will hold enough points, which are the mappings of the $\delta $ points in $\cal M$ to $\cal N$ inside the $\epsilon$, to reject the null at the 0.98 level or equivalently that the number of points mapped into the $\epsilon$ set has a probability less than or equal to 0.02, a common criteria. We choose $\delta$ to be large enough to produce the number of $\epsilon$ points to reject the null. If the number of $\epsilon$ points are not enough we enlarge $\delta$ and try again.

The questions are, what is a good null to choose and what is the probability distribution we should use? We propose a geometric null that answers these questions.

Once we choose a $\delta$ and gather the $n_{\delta}$ points it holds we assume some test value of $\epsilon$.  The $\epsilon$ is the radius of a sphere in $\cal N$. We now make the hypothesis that the $\delta$ points will fall at random into a sphere centered on $\phi ({\bf x})$ that is twice the volume of the $\epsilon$-sized sphere. Then the probability $p_{\epsilon}$ that each of the $\delta$ points will fall in the $\epsilon$-sphere is 0.5. The probability that, say, $n_{\epsilon}$ of those points will fall into the $\epsilon$-sized sphere is given by the cumulative binomial distribution. If this latter probabilty is below 0.02, then we reject the null that the points fall into the double-volume set by chance.  The natural steps here are to continue to shrink $\epsilon$ and $\delta$, as necessary, until the number of points are below a threshold where we cannot reject the null. 

Note that the value of $\delta$ matters only in that we use it to gather local points. The real measure of continuity once we assume a value for the the probability $p_\epsilon$ of each point landing in the $\epsilon$ set is the $\epsilon$ size and the probability of getting $n_{\epsilon}$ points out of $n_{\delta}$ points into the $n_{\epsilon}$ set. 

By choosing a $p_\epsilon$ for the null, then we can determined how many points one needs in the $\epsilon$ set to reject the null hypothesis given the number of nearby points in the domain. Let $n_\delta$ be the number of $\delta$ points in the domain, $n_\epsilon$ be the number of range time-twin points that fall in the $\epsilon$ set. Then the probability of getting $n_\epsilon-1$ or fewer points is given by the cumulative binomial distribution \cite{RiceJA, Kreys1} $C(n_\epsilon-1, n_\delta, p_\epsilon)$. A more realistic version of the probability would be to use the binomial distribution but for the situation where points are chosen without replacement.  This would involve hypergeometric binomial distributions and cumulative distributions\cite{RiceJA, Kreys1}. We found that with the small number of local $\delta$ and $\epsilon$ points we will deal with the more refined and complex binomial distribution without replacement didn't make much difference, so we use the simpler binomial cumulative distribution from here on given by,

\begin{equation}
  \label{eq:CumBinDist}
  C(n_\epsilon-1, n_\delta, p_\epsilon)=\sum_{j=0}^{n_\epsilon-1} \binom{n_\delta}{j}
  	p_\epsilon^j (1-p_\epsilon)^{n_\delta-j},
\end{equation}
which is the probability of getting $n_\epsilon-1$ or fewer points from a total collection of $n_\delta$ points in the $\epsilon$ set.

Then the probability of getting $n_\epsilon$ or more of the $n_\delta$ points in the $\epsilon$ set by chance is given by,
\begin{equation}
  \label{eq:ContPrb}
  P_>(n_\epsilon, n_\delta, p_\epsilon)=1-C(n_\epsilon-1, n_\delta, p_\epsilon),
\end{equation}
If the probability in Equation (\ref{eq:ContPrb}) is less than 0.02, then we reject the null and assume that $n_\epsilon$ points is sufficient to declare a local (at ${\bf x}_0$) continuous function from $\cal M$ to $\cal N$ at least to a local scale $=\epsilon$. Roughly speaking, the 0.02 value is two standard deviations into the tail of the distribution.

Table 1 shows the number of points from the $\delta$ set that have to land in the $\epsilon$ set to reject the null, i.e. accept that there is evidence at those numbers and set sizes to say there is a continuous function at ${\bf x}_0$. 

Generally, one tries to decrease the delta set to get the least number of points necessary to reject the null at the 0.02 level leading to the smallest set sizes where evidence for continuity can be claimed, which is 5 (see Table 1.). But note that there is not necessarily a linear relationship between the $\delta$ point distances in the domain and the distances of the time-twin points to the fiducial point in the range. One reason for this is because continuity does not require differentiability.  A fractal-like function can have a sampling of points close to the $\delta$ center that are not relatively close to the $\epsilon$ center. And even for differentiable functions on multi-dimensional objects, the distances can be stretched, shrunken, or distorted in different directions changing the order of the points distances compared to the order in the domain. 

The solution to this at each fiducial point is to keep track of each $\delta$-$\epsilon$ pair and choose the pair which rejects the null at the smallest $\epsilon$ value.
\begin{table}
\caption{\label{tab:table1}Number of $\delta$ points in domain and number of $\epsilon$ points necessary to reject the null hypothesis \cite{Pecora-attrRecon:2007}.  }
\begin{ruledtabular}
\begin{tabular}{lcr}
Num.$\delta$ pts. &Num.$\epsilon$ pts. &Num.pts.not in $\epsilon$\\
\hline
5 & 5 & 0\\
6 & 6 & 0\\
7 & 7 & 0\\
8 & 7 & 1\\
9 & 8 & 1\\
10 & 9 & 1\\
11 & 9 & 2\\
12 & 9 & 3\\
13 & 10 & 3\\
\end{tabular}
\end{ruledtabular}
\end{table}
A step further toward a continuity statistic is to calculate the average $\epsilon$ continuity value for many fiducial points on the data sets. We call this average $\epsilon^*$. The number of fiducial points to use can be estimated by comparing the volume of the data set with the total volume of the $\epsilon$ values found. When the total $\epsilon$ volumes are a substantial part of the volume of the data set, which can be fractal, then covering the set with more fiducial points would probably not change the statistic much.

One question remains: after we arrive at the value of the average continuity minimum $\epsilon^*$:  How can we turn the  $\epsilon^*$ value into a statistic? After all, there will always be an $\epsilon$ large enough to encompass enough $\delta$ points to reject the null, but how do we gauge that the $\epsilon^*$ is small enough to declare that there is evidence for a continuous function from $\cal M$ to $\cal N$?

There are at least two ways to construct a reasonable statistic and those are to compare $\epsilon^*$ to two sizes, which would rescale $\epsilon^*$ to sizes that are reasonably comparable with the scales of the data cloud. These two sizes are the average distance between points on the data cloud, say $\epsilon_{std}$ (given by the standard deviation of the data cloud) and the average of the closest distances between points on the data cloud, say $\epsilon_{min}$. 

Both of these statistics can be instructive, but generally $\epsilon_{min}$ is more informative since adding points to the data cloud will eventually decrease $\epsilon_{min}$ and $\epsilon^*$ should scale with $\epsilon_{min}$ if the function is truly continuous.  The alternate $\epsilon_{std}$, will not change and $\epsilon_{min}$ should decrease relative to its value. We will consider both ratios $\epsilon^*/ \epsilon_{min}$ and $\epsilon^* / \epsilon_{std}$ as continuity statistics, which are best used with varying numbers of total points in the drive and RC time series. 

Note that some thought about these statistics also implies that one cannot prove numerically that a mapping between data clouds is continuous. One can only say that there is evidence for continuity down to a scale $\epsilon^*$. There may exist discontinuities below that scale. Also, if the discontinuities are not dense in some way on the $\cal N$ data, then it may be easy for the numerical data to miss some discontinuities completely.  We have not considered these issues in more depth yet since we are mostly comparing data sets that are sure to have many discontinuities if there is no good embedding, but they are important to keep in mind.

There is one caveat to mention and it is important for finding nearest neighbor points in time series. In our search for points near the fiducial point we must avoid including points that are nearby in time since these will generally be correlated to our fiducial point at short times along the trajectory that included the fiducial point. In fact, when we find a "true" nearby point, i.e. one not near the fiducial point in time, we must then avoid temporally nearby points in time to that point, too. Nearby points in time can be added arbitrarily by simply sampling time series at a faster rate or calculating trajectories for smaller time steps, but this is an artificial increase of the topologically nearby points and will yield erroneous statistical conclusions. This is essentially saying that we need each point to come uniquely from a section of a nearby trajectory that is "temporally remote" from the point we just found. This is the concept of a "Theiler window" introduced by Ref[\cite{Theiler:1986aa}]. This principle is also carried through in the work by Kennel \cite{Kennel:2002aa} that introduced a similar concept called strands of points, i.e. sections of the trajectory associated with a point that is nearby the fiducial point.

In our numerical tests we use strands that are on the order of the number of time steps needed to decrease the linear correlation substantially.

\section{\label{sec:level4} Differentiability Statistic}

Differentiability is much less subtle than continuity. Basically, the idea is that a function is differentiable at a point if it can be well approximated by a linear function near the same point.  More technically, a general defintion in differential topology\cite{Chillingworth-DiffTop} is as follows. 

If $\cal M$ and $\cal N$ are vector spaces each with a metric ($d_{\cal M}$ and $d_{\cal N}$), respectively, then we can define a {\em norm} $\lVert \cdot \rVert$ on each that defines the "size" of a vector $\bf x$ by $\lVert {\bf x} \rVert=d_{\cal M}({\bf x},{\bf 0})$, where {\bf 0} is the origin of ${\cal M}$ and similarly for the vector space ${\cal N}$. Here we assume the metric is translationally invariant so we can define the norm using the metric. Then for a continuous function ${\bf \phi}$ that maps points in ${\cal M}$ to points in ${\cal N}$ we say that ${\phi}$ is {\em differentiable} at ${\bf x}\in {\cal M}$ if there is a linear map $D(\phi): {\cal M} \rightarrow {\cal N}$ such that,
\begin{equation}
\begin{aligned}
  \label{eq:DiffDef1}
  \forall \thickspace  {\bf h} \in {\cal M} & \text{ we have} \\
  & {\phi}({\bf x}+{\bf h})-{\phi}({\bf x})= D({\phi}) \boldsymbol{\cdot} {\bf h} +\lVert {\bf h} \rVert \eta ({\bf h}),\\
  & \hspace{23 mm}  = D{\phi} \boldsymbol{\cdot} {\bf h} +\lVert {\bf h} \rVert \eta ({\bf h}),
\end{aligned}
\end{equation}
where $\eta ({\bf h}) \in \cal N$ and $\lVert \eta ({\bf h}) \rVert \rightarrow 0$ as $\lVert {\bf h} \rVert \rightarrow 0$. We often write $D\,{\phi}$ in place of $D\,({\phi})$. Put simply, ${\phi}({\bf x})+D\,{\phi}\boldsymbol{\cdot} {\bf h}$ approximates ${\phi}({\bf x}+{\bf h})$. Since we are examining attractors on embedded manifolds we can restrict the above to a local coordinate patch on the manifold. For simplicity of notation we keep the local coordinate mapping implicit.

The manifolds holding the attractors are embedded in euclidean spaces, we need to add one more step and this will lead to a differentiability statistic.  The manifolds containing the attractors  usually have an intrinsic dimension smaller than the embedding space. They are like a "surface" in a higher dimensional space.  

If they are embeddings then their intrinsic dimension will be lower than or, at most, equal to the dimension of the embedding space.  Hence the differential of the mapping of the manifold will act only on the tangent spaces to the manifold. That is, if the object is originally a $k$-dimensional manifold and it is embedded in another space, it should remain a $k$-dimensional manifold and the derivative in the embedded space will act only on the tangent space to the embedded manifold.

This means, if ${\phi}$ is a diffeomorphism, the local linear maps (like $D\,{\phi}$) must span a dimension equal to the local dimension of the manifold containing the attractor. More mathematically if $\Delta$ is the (possibly fractal) attractor dimension locally at that point then we expect the local manifold dimension to be at least $\lceil {\Delta} \rceil$, where $\lceil {\bullet} \rceil$ is the ceiling function. And if there is a diffeomorphism ${\phi}$ between the embedded attractors the local tangent space dimensions of the drive and RC should be the same at ${\bf x}_0$ and ${\phi}({\bf x}_0)$. If the dimensions are the same, then we have evidence that we have a locally linear map between the drive and the RC. This will give us a statistic to help determine is we have a diffeomorphism, but it is strictly geometric and not (yet) a final answer.

We can calculate the local tangent space dimension by using a set of points on the drive and their time-twins on the RC to calculate the singular values (SV) of the set of local vectors. If the number of significant SV's are the same we know the local tangent spaces are of equal dimension. Note this only guarantees dimension equality, and does not fully involve the potential mapping $\phi$ between the drive and RC.  So we take the SV analyses further and analyze the functional relationship between the drive and RC attractors.

We start with Eq.(\ref{eq:DiffDef1}) and extend it to the inverse function $\phi^{-1}$ to help us test when $\phi$ and $\phi^{-1}$ are truly inverses of each other on tangent spaces at various drive and (time-twin) RC points.

The form of a differentiable inverse of Eq.(\ref{eq:DiffDef1}) is,    
\begin{equation}
\begin{aligned}
  \label{eq:DiffDef2}
  \forall \thickspace  {\bf h} \in {\cal M} & \text{ we have} \\
  {\phi}^{-1}\circ {\phi}({\bf x}+{\bf h})= 
  & {\bf x}+D{\phi}^{-1}{\cdot}D{\phi}{\cdot}{\bf h}+ \\
  &D{\phi}^{-1}{\cdot}\lVert {\bf h} \rVert \eta (\bf h)+\lVert {\bf u} \rVert \nu({\bf u}),
\end{aligned}
\end{equation}
where ${\bf u}=D{\phi}\boldsymbol{\cdot} {\bf h}+ \lVert {\bf h} \rVert \eta (\bf h)$
and $\nu({\bf u})$ is a function analogous to $\eta$ that goes to zero along with its argument. In Eq.(\ref{eq:DiffDef2}) the 3rd and 4th terms approach zero faster than the 1st and 2nd terms, leaving us with,
\begin{equation}
  \label{eq:DiffDef3}
  {\phi}^{-1}\circ {\phi}({\bf x}+{\bf h})={\bf x}+{\bf h}\approx {\bf x}+ D{\phi}^{-1}{\cdot} D{\phi}{\cdot}\bf h
\end{equation}
We see that to maintain the approximate equality for all small $\bf{h}$ we require $D{\phi}^{-1}{\cdot} D{\phi}=\bf 1$, where $\bf 1$ is the identity matrix. Thus, a diffeomorphism generates a tangent space mapping that is inverse to the tangent space mapping of the inverse diffeomorphim in each tangent space in the attractor's manifold.

We now have two statistics that combine to give us a complete statistic to test a homeomorphism for differentiability and whether we can claim that it has more structure to it and so is a diffeomorphism. The dimension of the tangent spaces in the drive must be the same at each point and their images in the RC must have the same dimension. Then on tangent spaces at time-twin points we can test at each time-twin points pair in the drive-RC system if the tangent space maps and their supposed inverses are indeed matrix inverses of each other.

There are some important considerations we must point out for the diffeomorphism statistic calculation.  We gather points in a neighborhood large enough to calculate the dimension locally. For the singular value decompositions (SVD) \cite{Golub_VanLoan_Mtx} in the drive and RC space we want the number of points we gather (not counting the fiducial point) to at least be 1 more than the number of meaningful singular values. Note that the number of points to claim continuity does not have relevance here. At this point we suggest we get enough nearest neighbor points around the fiducial point to be at least double the number of relevant singular values.

Let $K$ be the dimension of the drive space and $NM$ be the dimension of the RC space. $N$ is the number of nodes and $M$ is the number of dynamical variables of each. Let $Q^d$ be the $L \times K$ matrix of $L$ local vectors around a fiducial point ${\bf x}_0$ in the drive. Let $Q^r$ be the $L \times NM$ matrix of local vectors around a fiducial point ${\bf r}_0$ in the RC. For reasons we explain below we choose $L$ to be an even number $\ge$ the number of singular values of $Q^r$ and $Q^d$.

We first check that the number of singular values $P^d$ of $Q^d$ equals the number of singular values $P^r$ of $Q^r$. If not, then there cannot be a diffeomorphism as mentioned above. If $P^d=P^r$ we write $P$ without superscripts for the local dimension of the drive and RC and we calculate the local linear maps between the drive and RC and check that they are indeed inverses of each other. To do this we pick at random 1/2 of the $L$ points and calculate the "forward" (drive $\rightarrow$ RC) $D\phi$ matrix. Then we use the remaining 1/2 of the $L$ points to calculate the "inverse" (RC $\rightarrow$ drive) $D\phi^{-1}$ matrix. We use different sets of points for each calculation to avoid the trivial result that the calculated $D(\phi)$ matrix would have an inverse, $D(\phi^{-1})$  since the points are merely "reversed", i.e. the domain and range of the points are exchanged.  This explains why we choose $L$ to be at least twice $P$, the dimension of the drive. For better statistical sampling we also choose $L$ to be even larger.

We then examine the product $D\phi^{-1}{\cdot}D\phi$. If this is approximately equal to $\bf 1$ to some small deviation, then we conclude (statistically) that the mapping between the drive at the current fiducial point is a diffeomorphism.  We do this for several points on the attractors and examine the average statistic. The question of how to judge the choice of singular value cutoffs and allowed error in the $D\phi^{-1}{\cdot}D\phi$ will be addressed in the example cases later.

What remains is augmenting the above analysis with a null hypothesis as we did with the continuity statistic.  This means asking if the result of the product of the linear tangent maps being near to $\bf 1$ could happen "accidentally."  At this point we have not yet developed such a null test, so we will rely on the statistic being much less than 1.0 as evidence of a diffeomorphism.

\section{\label{sec:level5} One directional differentiability}

Here we show a statistic for testing functional differentiability in one direction only. We will not use this in the reservoir examples in this paper, but we show an example of it in the supplemental material about the statistics of continuous, but not differentiable functions.

To develop this statistic we borrow from the diffeomorphism statistic. We assume that we have a function $\phi$ which is differentiable in only one direction (from domain to range) as defined in Eq.(\ref{eq:DiffDef1}). We use SVD to find the dimension $P$ of the tangent space at a fiducial point $\bf x$ and the dimension of the tangent space in the range at the image of $\bf x$. As before, we find an even number of points ($L>P$) at a fiducial point $\bf x$.  We use $L/2$ points to find a local matrix approximation for $D\phi$. 

The system of points from the domain (the drive here) to the range (the RC here) has to be equal- or over-determined\cite{Golub_VanLoan_Mtx}. This is not to just deal with possible noise in the measured time series (since we are considering that the time series may come from measurements), but also that we cannot have a system that is underdetermined, but has a local linear map from the domain to the range. That is, the dimension of the manifold in the domain must be less than or equal to the dimension in the range. The SVDs of the local domain and range manifolds test for this and for determining the form (dimensionality) of the local derivative.

We get the approximate $D\phi$ using the usual least squares approach \cite{Golub_VanLoan_Mtx} to map the remaining points to the range of the function and compare these vectors to the actual vectors in the range using root mean square average differences, expressed as a fraction of the actual vector magnitudes. Small values of these fractions suggest that the function from domain to range is a differentiable function. We judge the function's differentiability on the domain by averaging over the fractional errors in the vectors. As with the differentiability statistic we would eventually like to develop a null hypothesis to aid in understanding the meaning of the statistic better.

\section{\label{sec:level5} An important point about functional relations between two coupled attractors}

In the following section we will use the symbols $\mathcal D$ for the drive attractor and $\mathcal R$ for the RC attractor.

We note first that we must test the statistics in two "directions": (1) a mapping from $\mathcal D$ to $\mathcal R$ and (2) a mapping from $\mathcal R$ to $\mathcal D$. This guarantees that the inverse of the "function" from the drive to the RC is invertible on a point by point basis.  If there are no or very small discontinuities, this would appear to be mostly guaranteed for dynamics that are calculated numerically or measured since the points are temporally paired along continuous trajectories. 

\begin{figure}
\includegraphics[ width=\columnwidth,height=0.3\textheight]{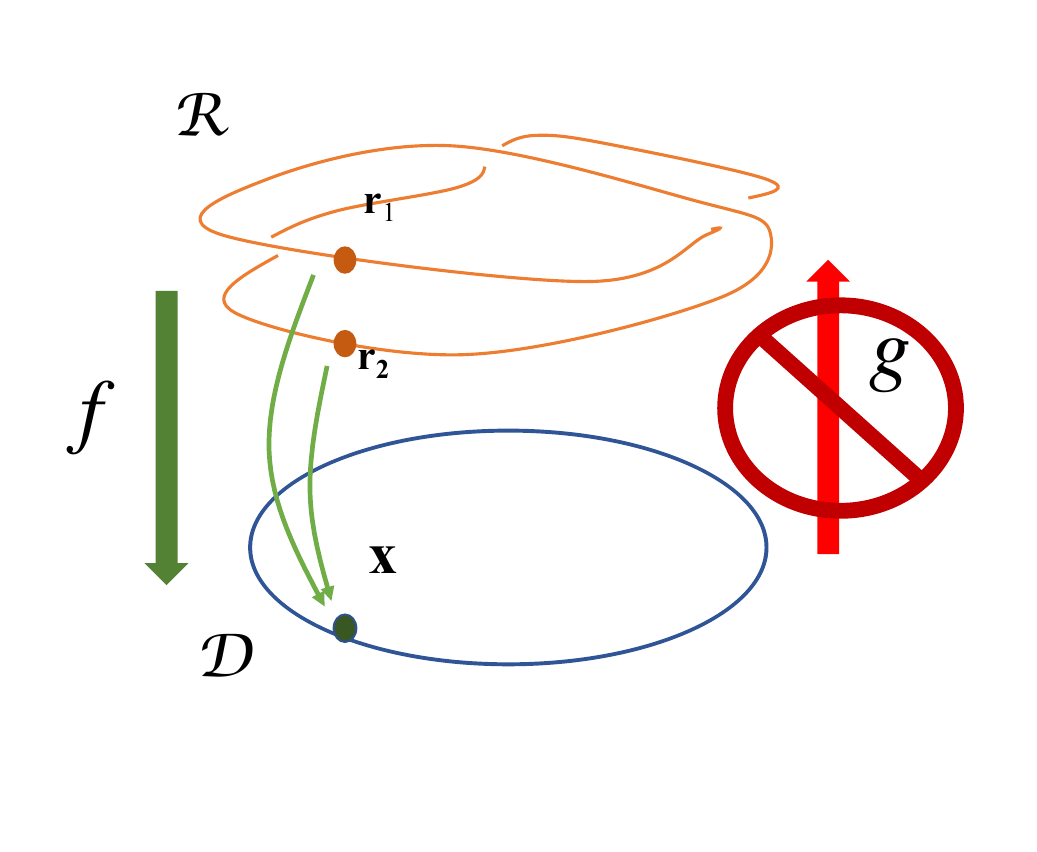}
\caption{\label{fig:PerdDbld_fig} Period Doubled drive-RC system. The drive is a periodic system and the RC is a period doubled response system. There is a function that maps each point in the RC trajectory to a point in the drive system, but the inverse "function" (drive $\rightarrow  $ RC ) cannot exist since there are two points in the RC corresponding to one domain point in the drive.}
\end{figure}

However, it is possible that the dynamics still do not yield invertible maps between the drive and the RC. A simple example of this is in period doubled systems where the drive is a periodic system and the RC is period doubled, quadrupled, or the like. This situation is shown in  Fig. \ref{fig:PerdDbld_fig}. Each data point in $\mathcal D$ will generally be associated to only one point in $\mathcal R$, but this is because of the temporal pairing and is misleading. It's easy to see that in the case of multiple period behavior there is a function $f: \mathcal R \rightarrow \mathcal D$. But there is not a function $g: \mathcal D \rightarrow \mathcal R$, since by definition a function cannot be one to many. Hence, continuity in both directions must be tested.

The solution to this situation is the continuity statistic itself. This is easy to see by assuming in Fig. \ref{fig:PerdDbld_fig} that we have a function $g: \mathcal D \rightarrow \mathcal R$. When we view $\mathcal D$ to be the domain of $g$ we search for small groups of points in $\mathcal R$ whose "partners" are in "local" sets of points.  We will likely get points in $\mathcal R$ that are in in both branches of the period-2 attractor since we will be using the "Theiler window"\cite{Theiler:1986aa}. This will lead to the detection of a "discontinuity" size that shows up in the continuity statistic approximately as the same $\epsilon^*$ at that attractor location no matter how many points we have in the data set. In this and any case where there are global "discontinuities" like this one we will not see the trends we see later in this paper where there is evidence for a continuous function and $\epsilon^*$ does change in a predictable way as the number of data points in the time series changes.

\section{\label{sec:level6} Attractor Comparison Statistic\protect}

There are situations where attractor properties are not as subtle as continuity or differentiability.  For example, when there are changes in initial conditions there is the chance that a dynamical system may settle into a different attractor. In other words, the initial conditions are in different basins of attraction and for reservoir computing the training of the first attractor might not apply to the testing using the second attractor. Yet, mostly because of the high dimension of the RC it is difficult and/or time consuming to find where the two attractors differ.  We exhibit such a situation in our examples and here show that there is a simple, but important statistic that can test for such a situation.

In this case we assume we have two time series that explore most of the attractors (labeled $\mathcal A$ and $\mathcal B$) for each, but are not sampled simultaneously. So the points in each are not time twins of each other, nor are they even of the same number (although the spatial dimensions must be the same). In this case we want a statistic that can distinguish whether two different time series are sampling the same or different attractors. To develop such a statistic we borrow concepts from set comparisons in metric spaces, which we modify to apply to data sampling of two or more sets rather than having access to all points in both sets, which would be impossible for sets containing a continuum of points.

An analytical definition of distance between two sets is given by the Hausdorff definition \cite{Rote1}

\begin{equation}
\label{eq:Hausdist}
d_H(\mathcal A,\mathcal B):=\text {max}\left\{ \sup_{a \in \mathcal A} \inf_{b \in \mathcal B}, \sup_{b \in \mathcal B} \inf_{a \in \mathcal A} \right\}
\end{equation}
This analytic form inspires our definition of set distances in data clouds, but while the form in Eq.(\ref {eq:Hausdist}) is necessary for mathematical proofs, the use of sup and inf is problematic for data analysis where a single point can distort the numerical analysis.  For this reason we use a definition of the average Hausdorff dimension similar to usage in some data analysis (see, for example, \cite{Aydin}.), but we keep the distances from $\mathcal A$ to $\mathcal B$ and $\mathcal B$ to $\mathcal A$ as separate statistics since one attractor can be arbitrarily small, but the other arbitrarily large. That is, the distances depend on the shape and placements of the sets with respect to each other and they are not necessarily even near equal. For example if $\mathcal A$ is small and placed near one side of $\mathcal B$, but $\mathcal B$ is large and has points distant from $\mathcal A$, then, on average, the distance from $\mathcal A$ to $\mathcal B$ will be small, but the distance from $\mathcal B$ to $\mathcal A$ will be large. 

So, we replace inf with min and we use two distances to test whether $\mathcal A$ and $\mathcal B$ are close:

\begin{equation}
\begin{aligned}
\label{eq:setdists1}
d_{avg}(\mathcal A,\mathcal B):=\text{avg}_{a \in \mathcal A}\: \min_{b\in \mathcal B}\: d(a,\mathcal B)\\
d_{avg}(\mathcal B,\mathcal A):=\text{avg}_{b \in \mathcal B}\: \min_{a\in \mathcal A}\: d(b,\mathcal A),
\end{aligned}
\end{equation}
where the order of $\mathcal A$ and $\mathcal B$ in the distance formula matters. We will see that these simple statistics are very helpful when we encounter multiple attractors in high-dimensional systems like an RC.

There is an important extension to the statistics in Eq.(\ref {eq:setdists1}) that we can add in the case of time series from dynamical systems. We made the point earlier that for many time series the points nearby to each other in time will be well correlated. We can use this correlation (easily measured) to increase the power of Eq.(\ref {eq:setdists1}) statistics.  Points in time into the past or future of the closest points will also be close up to some fraction of the correlation time length. Hence, we can find these closest points by searching (in time) along the trajectories of attractors $\mathcal A$ and $\mathcal B$ add those points' distances to the averages in Eq.(\ref {eq:setdists1}), thus using the dynamics to make them a more reliable gauge of attractor identity. This is similar to Kennel's use of "strands"\cite{Kennel:2002aa}. We do this in the following sections which use the attractor comparison statistic (ACS).


\section{\label{sec:level7} Numerical examples of diffeomorphic statistics\protect}

In these numerical studies we use as the drive system the Lorenz 63 ODE model along with an RC network with polynomial nodes which will be integrated to produce the dynamics.  

The RC network structure was 100 nodes randomly generated and scaled to have a spectral radius slightly less than 1.0. The inputs to the RC nodes were randomly scaled through input coefficients of values between -1.0 and +1.0. The RC was driven with the Lorenz $x$ component and the training of the reservoir was done to reconstruct the $y$ and $z$ components.  The trainings were also tested using the same Lorenz drives, but started at different initial conditions. 

With the exception of the ACS, the results presented here are all averages over 20 to 30 percent of the systems' attractors, as determined from the sizes of the local sets of points used to calculate the various statistics. The statistics were calculated with increasing numbers of points until the results leveled off. The ACS was calculated using sets that covered 70\% or more of the systems' attractors.  The standard deviations of all statistics were also calculated, but all showed no large variations in the the statistics' groups and, hence, are not presented here. 

The ODE equations for the Lorenz 63 dynamics are,

\begin{equation}
\begin{aligned}
\label{eq:LorenzODE}
\frac{dx}{dt}&= \sigma (y-x)\\
\frac{dy}{dt}&= -xz+\rho x-y\\
\frac{dz}{dt}&= xy-\beta z,
\end{aligned}
\end{equation}
where $\sigma=10.0, \rho=60.0$ and $\beta=8/3$.

The ODE equations for the polynomial node RC are \cite{Carrolltimeshifts},

\begin{equation}
\begin{aligned}
\label{eq:polyRC}
\frac{dr_i}{dt}&= \alpha \Big[ \kappa (p_1r_i+p_2r^2_i)+p_3r^3_i \\
  & + \sum_{j=1}^N A_{ij}r_j+w_ix(t)\Big] ,
\end{aligned}
\end{equation}
\noindent where $\alpha=1.0$, $p_1=-4.0, p_2= -0.871984$ and $p_3=0.52492$.

%
%

In Eq.(\ref {eq:polyRC}) $\kappa$ will be varied to change the dynamics, especially the RC damping. These variations will change the stability of the RC allowing us to examine the quality of the RCs reconstructions of the unseen Lorenz variables ($y$ and $z$) as stability varies up to the "edge of chaos" and compare those qualities simultaneously with the change in the continuity and differentiability statistic.

Training and testing time series were generated using the {\fontfamily{TXTT}\selectfont odeint} function in the {\fontfamily{TXTT}\selectfont scipy.integrate} library of python. The time steps for the Lorenz system were (in the original Lorenz parameters) 0.02. The time steps for the RC were (in the given parameters) 0.1. The time series were each 100,000 points and and the times scales led to local linear correlations that were relevant up to 20 steps.

\subsection{\label{sec:level8} Continuity Statistic Test\protect}

Training and testing of the reconstruction of the $y$ and $z$ components were done using the dynamics of Eqs.(\ref {eq:LorenzODE}) and \ref {eq:polyRC}) for the case of $\kappa =1.0$. The average training errors for the reconstruction of the $z$ component (the harder of the two unseen components to reconstruct) were $4.0\times 10^{-3}$ for a 40,000 point time series, which is a highly accurate result. We extended this to calculating the training and testing errors of $z$ for different numbers of time series points $(1.0\times10^4$ to $1.0\times10^5$. The testing and training errors both decreased from $8.0\times 10^{-3}$ down to $1.0\times 10^{-3}$ corresponding to the increase in number of trajectory points. 

\begin{figure}
\includegraphics[ width=\columnwidth,height=0.5\textheight]{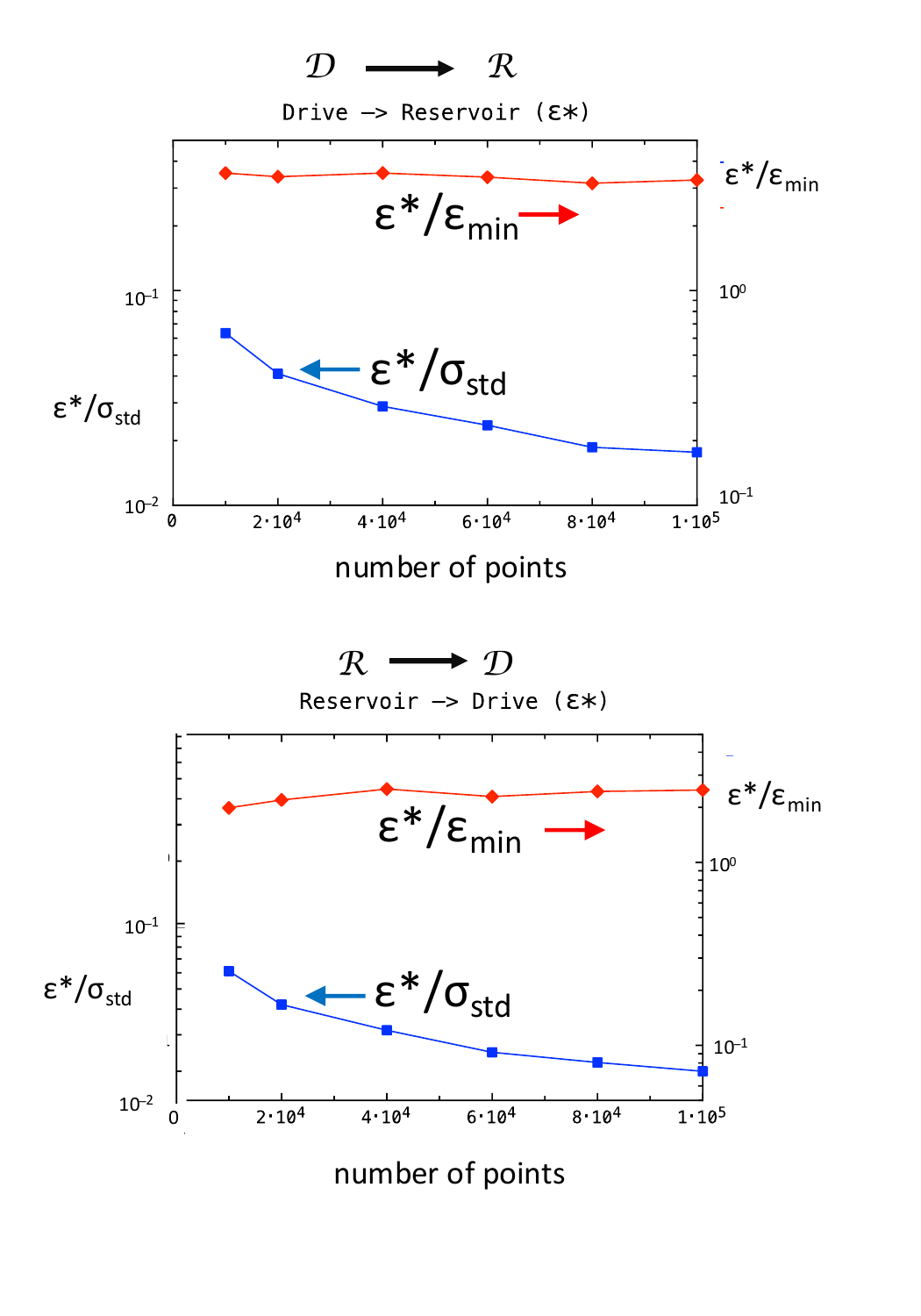}
\caption{\label{fig:epsstar_fig} Continuity statistic $\epsilon*$. Smaller $\epsilon^*$ values indicate better continuity. Hence, the ratio $\epsilon^*/\sigma_{std}$ should decrease as the number of points on the attractor increases (the density of points increases while the attractor size is constant). Since $\epsilon_{min}$ also decreases as we add attractor points, for good continuity statistics it makes sense that the ratio $\epsilon^*/\epsilon_{min}$ should remain approximately constant for a continuous fcn.}
\end{figure}

To examine how these reconstruction results compared to the continuity dependence on time series length we calculated both continuity statistics ($\epsilon*/\epsilon_{min}$ and $\epsilon*/\sigma_{std}$) for the same range of time series lengths. The results are shown in Fig. \ref{fig:epsstar_fig} for both statistics vs. trajectory lengths for each direction of potential mappings: from drive to RC and from RC to drive. From the figure we see that $\epsilon*/\sigma_{std}$ decreases in both functional directions as we would expect for a continuous function when we more densely populate the attractors. And, importantly, the $\epsilon*/\epsilon_{min}$ values remain constant implying that the continuity statistic is always near the minimal average distance between data points, again, as we would expect for a continuous function. These results suggest that we have a continuous and invertible mapping between the drive and RC, i.e. evidence for a homeomorphism, which is the first step in testing for a RC reconstruction of the drive attractor. 

As we mentioned earlier the hyperparameter space for RCs is large. The parameters can be indirectly tuned for certain drive systems by altering the RC and seeing how well it reconstructs the trajectories (as well as other RC tasks). However, a more direct way to probe the parameters that does not depend on an additional layer of dynamical calculations is to calculate the continuity statistic of the mapping between drive and RC. We can tune the parameters to guarantee the best statistical homeomorphism.

\begin{figure}
\includegraphics[ width=\columnwidth,height=0.5\textheight]{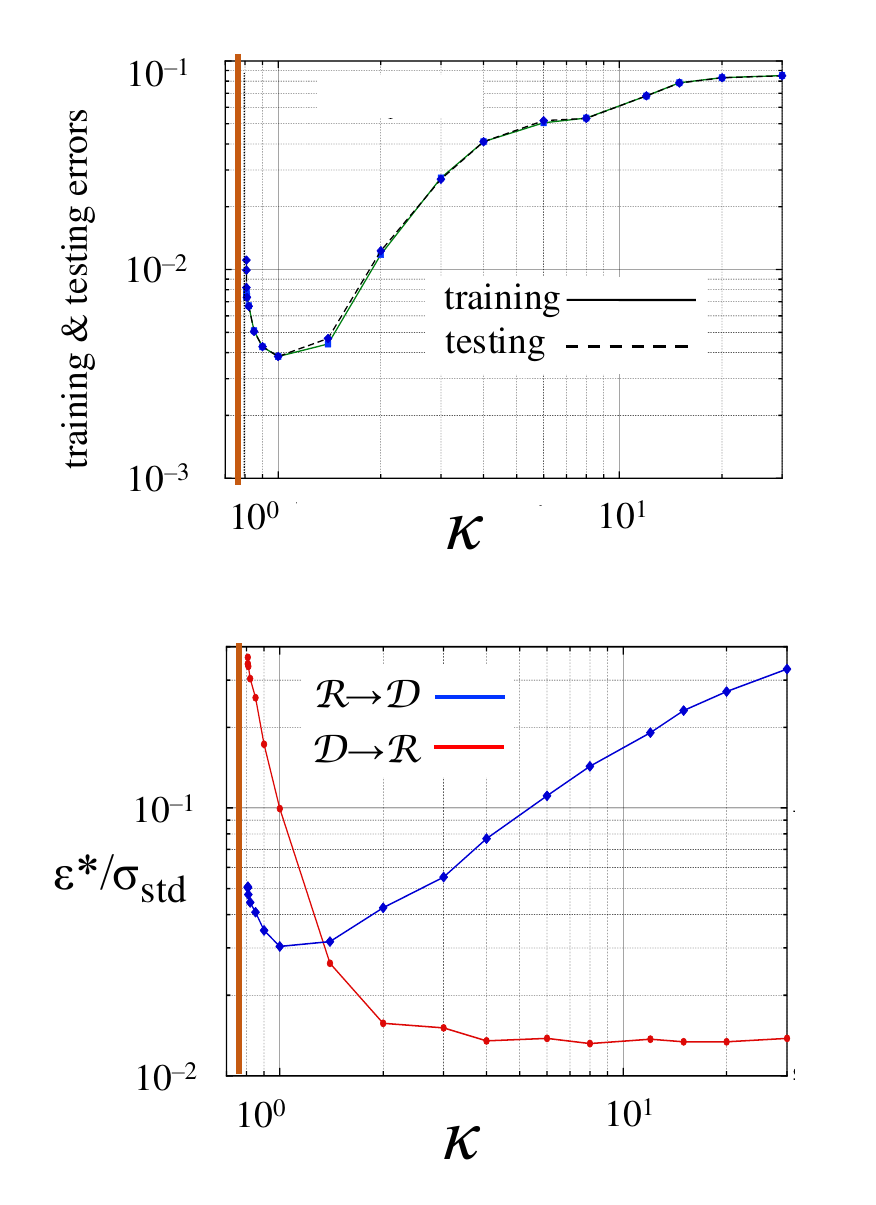}
\caption{\label{fig:Fig-Traing-Testg-and-CntStat} Training and testing errors and the Continuity statistic $\epsilon*$ as a function of the damping parameter $\kappa$. The vertical red line marks the $\kappa$ value where the RC becomes unstable, the so-called "edge of chaos."}
\end{figure}

We show how this would work using the current Lorenz-polynomial system and generating time series for various values of the damping parameter $\kappa$ and then calculating the continuity statistic (the $\epsilon^*/\sigma_{std}$ version) in both directions for the system. This is shown in Fig. \ref{fig:Fig-Traing-Testg-and-CntStat}.

The continuity statistic shows that the best mapping occurs for $\kappa$ between values of 1.0 and 1.5. This agrees well with the testing and training errors. Although the drive to RC continuity statistic becomes minimal above $\kappa=2$ the optimal bidirectional continuity statistic is in the range of the best testing errors.  Thus, we don't need to do an additional test reconstruction to adjust the $\kappa$ parameter.

The results in Fig. \ref{fig:Fig-Traing-Testg-and-CntStat} show that the training and testing errors increase as $\kappa$ approaches the value where the RC goes unstable, the vertical red line in the figure. This degrading of the continuity statistic (resulting in a poorer training and testing errors) is related to a dynamical phenomenon that is well-known in the dynamics field, but with only a few exceptions (for example, Ref.\cite{Carroll-Edge_ofChs}) has been noted in the work on RC systems. This is the phenomenon of weak vs. strong generalized synchronization \cite{Pyragas-WeakStrongSync,Rulkov-GenSync,Pecora-ChtDrvRespGenSync} or excessive filtering of chaotic signals\cite{Badii-FilterChsDim}. We mentioned this earlier in the claim of optimal operation of the RC at the "edge of chaos."

Weak generalized synchronization happens when the damping or dissipation of a system driven by a chaotic signal is small resulting in a long tail of mixing values of the driving system too far into the past. Because of the stretching and folding of the chaotic dynamics this causes points in the distant past to be "mixed" with more current points\cite{Kantz_Obrich-1997,Pecora-attrRecon:2007}. This changes the mapping of the drive from a smooth one to a fractal one and eventually gives large $\epsilon^*$ values. This is the reason for the so-called weak synchronization phenomenon, which results in an increase in the training and testing errors because of a loss of continuity resolution as seen in Fig. \ref {fig:Fig-Traing-Testg-and-CntStat}. 

A more mathematical explanation of this phenomenon is the Kaplan-Yorke conjecture \cite{KaplanYork-FractalDim}. This uses the Lyapunov exponents of the drive-response system (drive-RC in our case) to explain the change in dimension that can occur when the Lyapunov exponents of the drive overlap those of the response (RC). The formula for the (Kaplan-Yorke) dimension of the response is,

\begin{equation}
\label{eq:KapYorke}
D_{KY}= j+\sum_{k=1}^j\frac{\lambda_k}{\vert \lambda_{j+1}\vert},
\end{equation}
where $j$ is the index of the last Lyapunov exponent for which the sum is positive. 

Eq.(\ref {eq:KapYorke}) captures quantitatively the idea that if the dissipation of the driven system (the RC in our case) is not negative enough compared to the drive's exponents, the dynamics of the chaotic drive will add to the fractal structure of the response. Of course, if the RC has too much damping the past values of the drive will not affect the RC's behavior. This situation is described in more detail in the paper by Carroll\cite{Carroll-Edge_ofChs}. The situation for simple chaotic systems driving other dynamical systems (e.g. Lorenz driving a Rössler system) is covered in the earlier paper by Pecora and Carroll\cite{Pecora-ChtDrvRespGenSync} .

Both extremes of dissipation relate directly to our former explanation of how the drive's attractor is reconstructed in the RC and explain why too little or too much RC dissipation leads to poor performance of the RC. It is not that a fractal structure cannot be continuous image of another manifold, sometimes it is. The Weierstrass function is one example (see Supplement A).  But the amount of data necessary to overcome the "rough" fractal structure increases dramatically, hence the poor RC performance.  

Note that the differentiable nature of the drive-RC system will cease to be true in this case (as we will show below). No matter how much data we have or how the time series are sampled, a fractal structure is not a locally smooth manifold. This means the RC will not replicate the dimension of the drive or the drive's Lyapunov exponents. We show an example of this in Supplement A.

\begin{figure}[h]
\includegraphics[ width=\columnwidth,height=0.5\textheight]{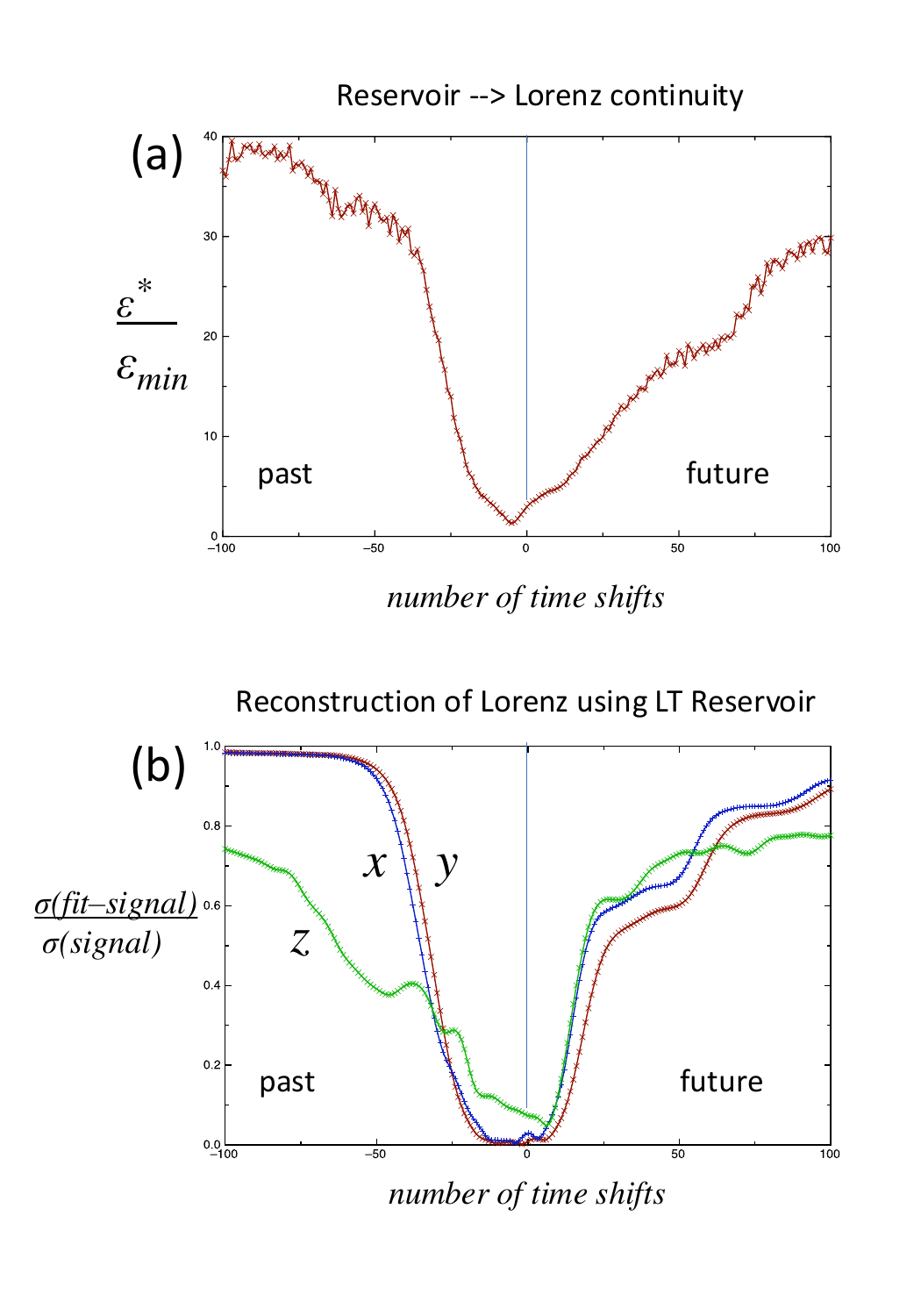}

Fig-TimeShift-Resv-to-Lor-cnt-fits/Fig-TimeShift-cnt-fit-R
esv to-Lor-ppt

\caption{\label{fig:TimeShift-Fig-Reservoir-to-Lor-fits} The (a) Continuity statistic (in terms of $\epsilon^*/\epsilon_{min}$) and (b) the training errors into past and future as a function of time shifts of the $x$ (blue), $y$ (red), and $z$ (green) Lorenz components. }
\end{figure}

\subsection{\label{sec:level9} Existence of continuous functions into the future and past\protect}

Until now we have examined the continuity statistic in both directions between the drive and RC, both at the same point in time. Given that the continuity statistic measures the likelihood of a continuous function from one data set to another (assuming points of the data sets are time twins), we can also ask other questions about other functions on the data sets.  One is, how good is a potential mapping from the drive to the RC at times in the past or future?  This will gauge how well past or future points in the drive can be reconstructed using the RC.

We ran training tests of the previously mentioned Lorenz-Polynomial drive-RC system but using offsets of the RC time series to train on the values of the drive from the past up to times in the future relative to the RC time points. We also calculated the continuity statistic of the RC to the drive over the same time shifts. 

In Fig. \ref{fig:TimeShift-Fig-Reservoir-to-Lor-fits}(a) the continuity statistic shows three basic features. One is that near small (around zero) time step shifts into the past the continuity (from the drive to the RC) improves for up to 10 steps, i.e. $\epsilon^*/\epsilon_{min}$ decreases. Two is that the continuity into the future decreases monotonically, i.e. $\epsilon^*/\epsilon_{min}$ increases. Three is that a function continuity into the distant (25 or more time steps) past becomes poor much faster than the statistic into the future by the same amount of time steps. These features show up in the training errors in Fig. \ref{fig:TimeShift-Fig-Reservoir-to-Lor-fits}(b). Some are more evident in each component whereas the continuity statistic in \ref{fig:TimeShift-Fig-Reservoir-to-Lor-fits}(a) in some sense is a mixture of all three.

With the examination of the continuity statistic between time series into the past or future, we can also think of the continuity statistic as a postdiction (into the past) and a prediction (into the future) test or, more accurately, a function from one time in a dynamical time series to another.  This helps us understand why the two plots in Fig. \ref{fig:TimeShift-Fig-Reservoir-to-Lor-fits} track each other. The good continuity into the future means we can successfully use the training to predict the values of the dynamical variables into the near future making it possible to use the RC as a autonomous dynamical system \cite{LuHunt, KongLai-RCBased}. 

The continuity statistic also shows that close points on the attractor at time shift=0 (the present) will be widely separated at times into the more distant future and past. The training will not be constrained since there will be a large divergence of nearby points into the past and the future leading to poor data fitting of the output weights.

\subsection{\label{sec:level13} Diffeomorphism Statistic\protect}

The calculation of a diffeomorphism statistics involves the (above) calculation of the continuity statistic and then the calculation of the (two-way) differentiability given previously that estimates at many points on the attractor the average product of the numerical derivative matrices $D\phi^{-1} D\phi$, where we remind that the matrices are each calculated separately from two different local sets of points. This product should be close to the identity matrix $\bf 1$. We use the following two-way derivative statistic (based on the Fibbonacci matrix norm) to calculate the average difference per component of $D\phi^{-1} D\phi$ from $\bf 1$:

\begin{equation}
\label{eq:DiffStatDef}
\sqrt{\sum_{i,j}\big( D\phi^{-1} D\phi-{\bf 1}\big)_{ij}^2/P},
\end{equation}
where $P$ is the local manifold dimension (see section \ref{sec:level4}).

Obviously, the smaller this derivative statistic is, the more we can claim that the mapping $\phi$ is a diffeomorphism.  This is a relative comparison and, as yet, we have not devised an absolute range or a null hypothesis in which we can say that the mapping is likely differentiable in both directions. It appears that as in the continuity statistic case, the statistic will depend on several variable things, like the number of data points used. However, for now we will use the differentiability statistic in a relative way to continue our analysis of the Lorenz-Polynomial system as in section \ref{sec:level3}.

\begin{figure}
\centering
\includegraphics[width=9cm]{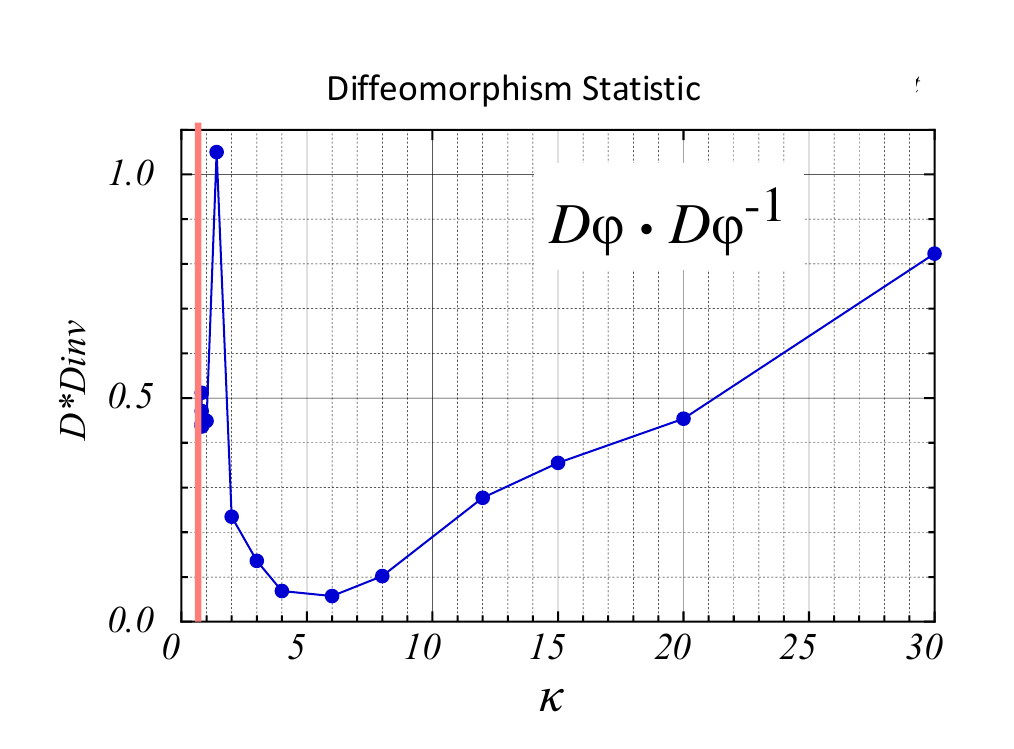}
\caption{\label{fig:Fig-DiffTest-RC-Lor} The Diffeomophism statistic as a function of damping parameter $\kappa$. }
\end{figure}

In Fig. \ref{fig:Fig-DiffTest-RC-Lor} we show the differentiability statistic as a function of the damping coefficient $\kappa$. Comparing this to the continuity statistic for the same $\kappa$ range we see that the minimum occurs at about $\kappa=5.0$.  This is slightly larger than the 1.0 to 2.0 range for the continuity statistic, but it shows the same behavior as $\kappa$ is increased from 1.0. This further suggests that the best embedding is between $\kappa=1.0$ to $5.0$, which agrees with the testing data. It appears that the continuity statistic is a better guide in this case, although the differentiability statistic provides more evidence for the existence of a minium and that near the instability threshold the differentiability, like the continuity statistic becomes worse. 

\subsection{\label{sec:level14} Attractor Comparison Statistics and RC basins of attraction\protect}

Dynamical systems can have multiple attractors for some set of parameters, which includes driven systems. This means that what attractor the RC can settle into in different dynamical patterns in separate regions of the state space will depend on the initial conditions of the entire system, the drive and the RC. This was shown to occur in RC systems in the paper by Flynn et al.\cite{Flynn-DoubleRC}. Interestingly, in our case since the RC starts from an equilibrium state, where all the nodes are at the origin of its state space (all RC variables, ${\bf r}_i=0$), but the final state depends on the drive system's initial state, even though the drive still ends up in the same state. This situation is where our Attractor Comparison Statistic (ACS) will be helpful.

We run into this problem of multiple attractors when we have a Lorenz drive and a slightly different polynomial node RC system that we are using to train on reproduction of the $y$ and $z$ components from an $x$-driven RC and then test that reconstruction using a drive started at a different initial condition. The RC node dynamics are given by the same ODE system in Eq.\ref{eq:polyRC} with the overall coefficient $\kappa=1.0$ and we vary the damping coefficient $p_1$ in the polymomial over the range $-7.0$ to $-0.5$.

\begin{figure}
\centering
\includegraphics[width=9cm]{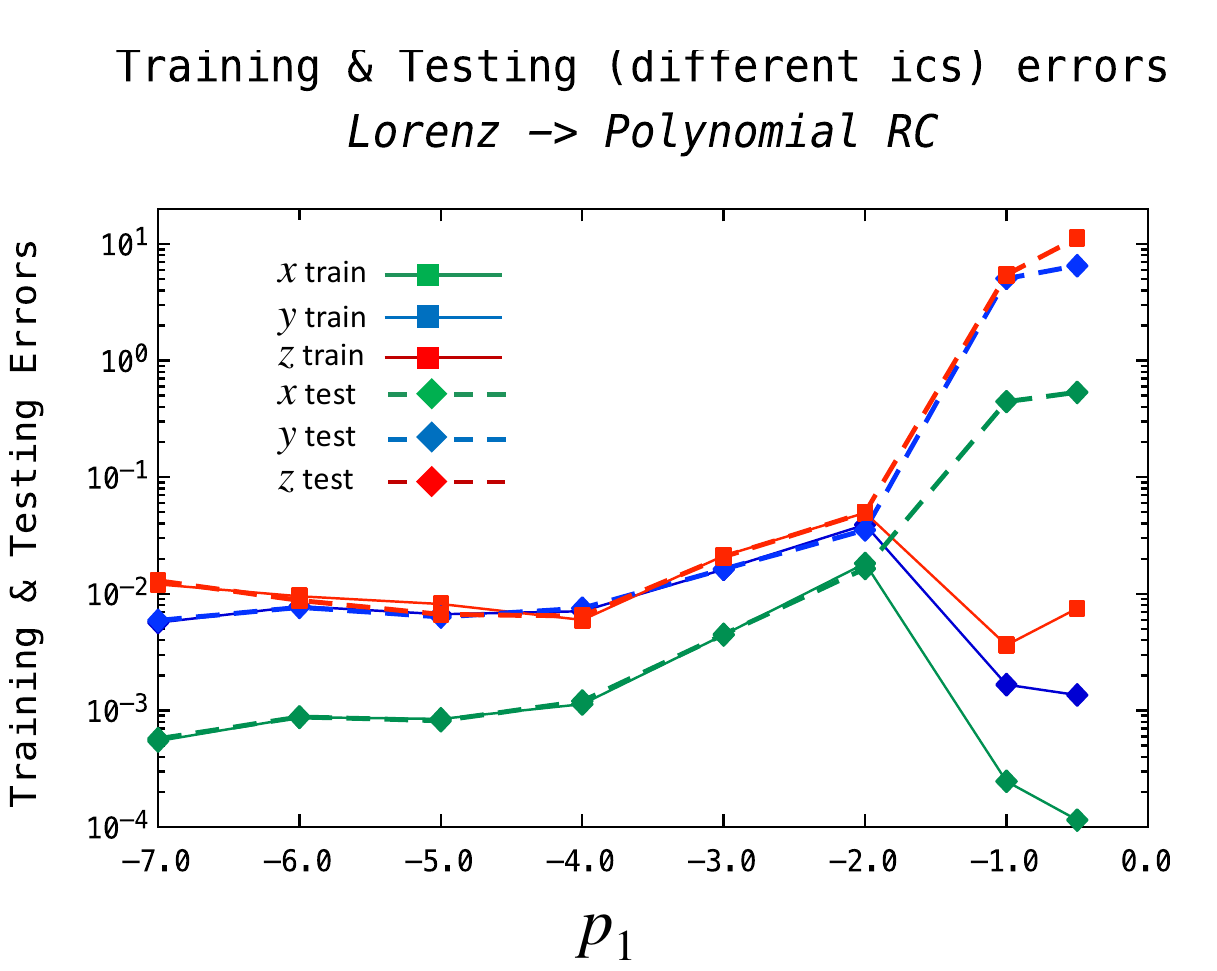}
\caption{\label{fig:Fig-TestTrain} Training and Testing errors as a function of damping parameter $p_1$. }
\end{figure}

In Fig.\ref{fig:Fig-TestTrain} we show the testing and training errors as we vary $p_1$. The training and testing errors are small and track each other well until we get to $p_1=2.0$ Here we see a change in the testing errors for all Lorenz components of more than a factor of $10^3$.

\begin{figure}
\centering
\includegraphics[width=9cm]{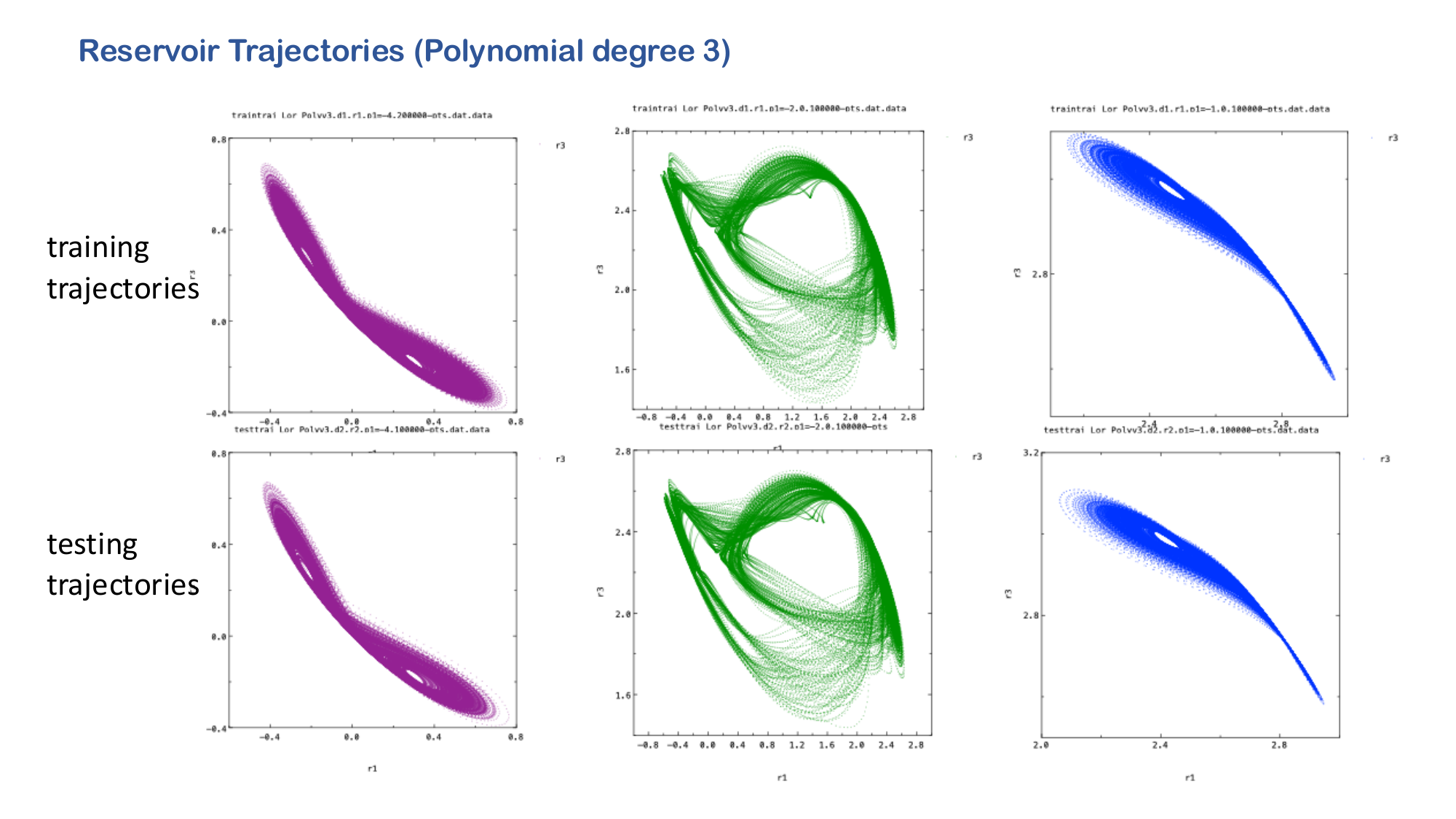}
\caption{\label{fig:Fig-Traing-Testg-trajs-atdfntp1} A sample of training and testing trajectories for $p_1$ below, at, and above where the testing errors diverge from the training errors. }
\end{figure}

If we examine some of the 100 RC nodes' dynamics in pairs of components at points below, near, and above the dramatic testing error range we see very similar attractor shapes (which are, of course, projections from a higher 100 dimensional state space).  These are shown in Fig.\ref{fig:Fig-Traing-Testg-trajs-atdfntp1}.

\begin{figure}
\centering
\includegraphics[width=9cm]{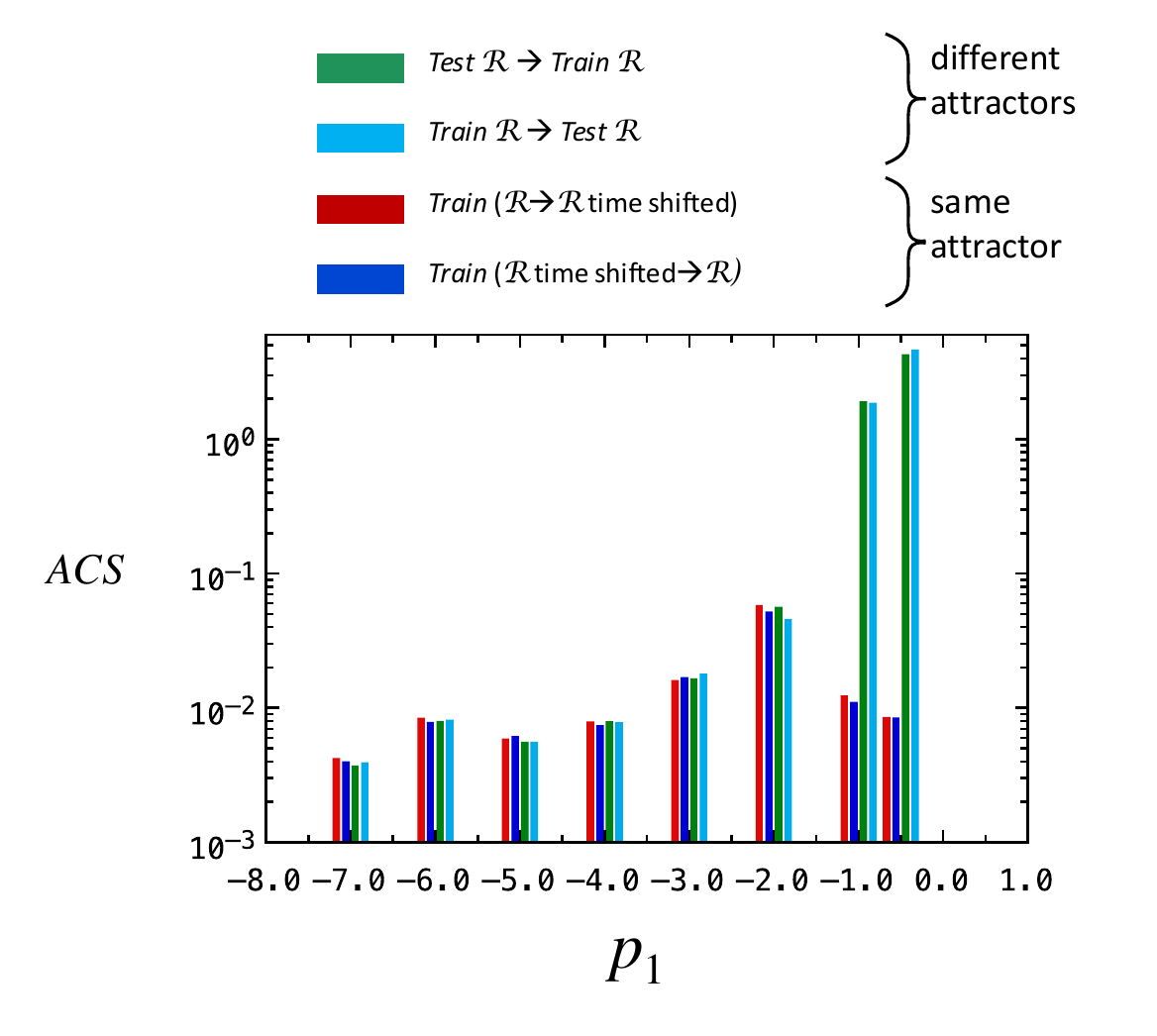}
\caption{\label{fig:Fig-ACS-train-test-RCAttrs-ACS} ACS values in both directions as a function of damping parameter $p_1$. }
\end{figure}

Rather than examining trajectory pairs for 100 nodes or even numerically comparing them, the simpler approach is to use the ACS. We tested the ACS by first computing the ACS between  portions of the $(x,y,z)$ points from the original training attractor consisting of time series from two, non-overlapping 20000 point segments, i.e. one is shifted more than 20000 in time.  Recall that the ACS is directional in that the statistic must be calculated in both directions between the attractors. 

The results of the ACS are shown in Fig. \ref{fig:Fig-ACS-train-test-RCAttrs-ACS}. The ACS is small (approximately $10^{-2}$ or smaller for $p_1$ values at $-2.0$ and below for train-(time-shifted) train and test-train systems. But for larger $p_1$ values ($> -2.0$) the ACS for the comparison of train to test increases to above $10^0$, i.e. 100 times the other ACS at more negative $p_1$ values. Meanwhile the train-(shifted) train ACS values remain small. Given the radical change in the trajectory of the components of each system's coordinates at $p_1=-2.0$, we suspect that near that parameter value the RC system is undergoing a bifurcation at that $p_1$ value. This implied that the RC attractors were different in this parameter range.

\begin{figure}
\centering
\includegraphics[width=9cm]{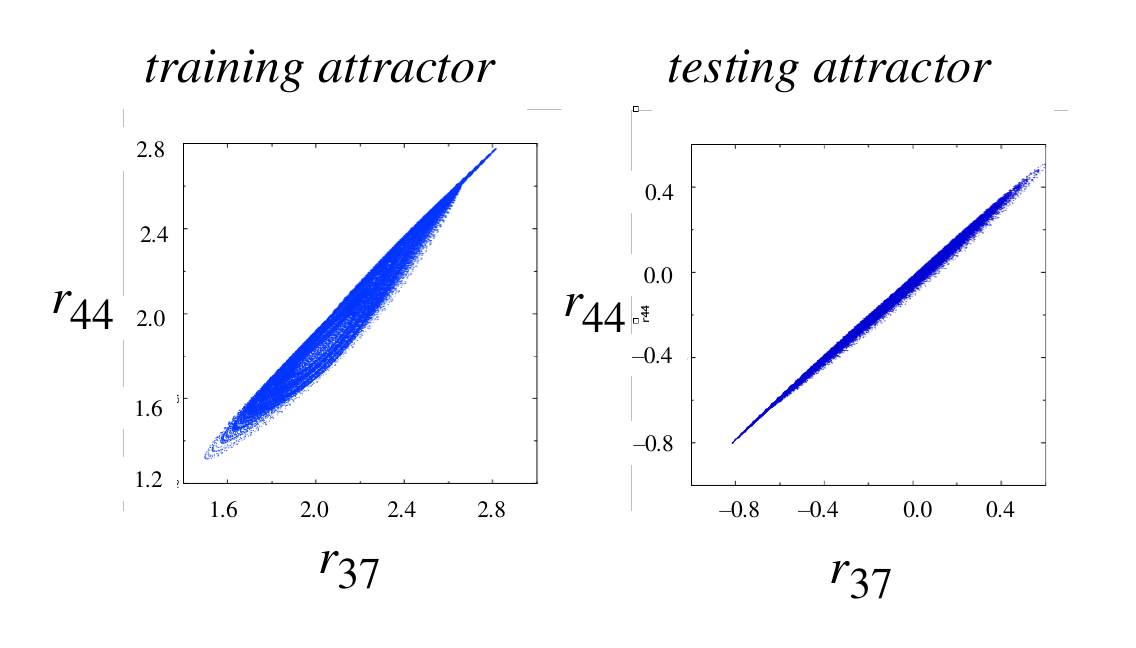}
\caption{\label{fig:Fig-RC-train-test-coords} Training and Testing errors as a function of damping parameter $p_1$. }
\end{figure}

In examining in detail the trajectories in detail in each RC coordinate.  We eventually found two coordinates in which the training and testing systems differed for $p_1 > -2.0$ from the systems below that value. All other coordinates were the same. These two are shown in Fig. \ref{fig:Fig-RC-train-test-coords}. In this parameter region the plots of the system look similar (both are thin and slanted the same way, although the testing attractor appears to be tilted at a smaller (acute) angle. 

However, the real difference is that the training attractor is centered in the $r_{37}-r_{44}$ plane in the region around the value of 2.0 and the testing attractor is centered in the $r_{37}-r_{44}$ plane in the region around the value of -0.2.  The test attractor is essentially very similar (almost identical) to the training attractor, but is shifted (translated) along the two axes shown by (approximately) $-2.2$, which is enough to keep the attractors in separate regions of state space and cause the large ACS changes shown in FIG.\ref{fig:Fig-ACS-train-test-RCAttrs-ACS} .

\section{\label{sec:level14} Conclusions\protect}

The continuity and differentiability statistics in both directions gives two pairs of statistics (one for $\mathcal D \rightarrow \mathcal R$ and another for $\mathcal R \rightarrow \mathcal D$) that can give evidence that we either have a diffeomorphism or not between the drive and RC. As we mention above, the statistics can be used in more general situations to test for the existence (or not) of functions between more generally (paired) data sets. In a sense the continuity statistic suggests a function exists that predicts one set of data points from knowledge of another data set of points.

We addressed the case here where we have access to all dynamical variables.  However, if we have a subset of variables that we can record, we can use a hybrid approach using delays in the recorded variables to reconstruct the state space trajectories and attractors. And, if the dimension of the drive is not too large, we will need to only reconstruct in a subspace of lower dimension to test for a diffeomorphism. Note that this approach may not always work. Our example using the ACS shows that sometimes a small number of RC nodes may have different dynamics from the trained dynamics.  Then, curiously, the Takens reconstruction may not expose those.  This situation needs to be studied in more detail.

We also showed that checking the differentiability statistic in both directions  can be done with the diffeomorphic test for a possible local linear map between two (paired) data sets. But the test for a one-way differentiable function from one data set to the other can also be easily done, too and used when there is only a continuous function in one direction (see our period doubled counter-example earlier in this paper). 

Gauging the size of the errors in all these statistics is something that needs more refinement, especially in the differentiability statistics.  The continuity statistic error relies on a sensible null hypothesis which at least gives a geometric understanding of the statistic.  The differentiability statistic still needs something more geometric related to a null hypothesis, which we are still working on.

These statistics can be useful in experiments or hardware implementations where a good model of the RC is not known, but time series can be recorded. In this situation we can’t calculate Lyapunov exponents, but we can still make statements about functional relations between two coupled dynamical systems.  

We also must mention a recent publication that focuses on the relation between a Takens embedding and a RC, focusing on RCs constructed with time-delay dynamics\cite{DuanKurths-RCEmbdgDelaySys}.  This appears to aid in RC embedding even up to the point of using a single time-delay system to reconstruct the drive attractor. Applying the diffeomorphism statistics would be interesting in this case and could help guide the development of the time-delay node(s). 

We finally note that the statistics we present here are very general, especially the continuity statistic. We assume almost nothing about the data sets or the geometric structures they possibly have.  We list these advantages as follows,

$\bullet$ We assume nothing about the possible functional relations between the data sets.

$\bullet$ The statistic is for one direction only ($\mathcal D \rightarrow \mathcal R$). It says nothing about the inverse. 

$\bullet$ The inverse is a separate independent statistic, ($\mathcal R \rightarrow \mathcal D$)

$\bullet$ The statistic is inherently local.

$\bullet$ The statistic is dependent on the number of points in the data set as would be expected for data.

$\bullet$ $\epsilon^*$  is approximately the relative size of the smallest discontinuity we can detect.

$\bullet$ If $\epsilon^*$ scales with and is near the size of $\epsilon_{min}$, then this is further evidence of a continuous function.

$\bullet$ This is a statistic giving evidence (or not) of a continuous function. It is not a mathematical proof.

$\bullet$ Derivative or diffeomorphism tests are useful in determining if fitting data or time series to some smooth functional basis will be worthwhile.

\section{\label{sec:level14} References}

\bibliography{NLDofRC_3}



\section*{Supplement A: Continuity and Differentiability of the Weierstrass (Fractal) Function.}
 The Weierstrass function is a formula for approximating a real continuous, but non-differentiable function on the real line with a type of Fourier series as follows\cite{Mandel-FractalGeom, WeiersFcn-Wiki}:

\begin{equation}
\label{eq:WeiersFcn}
w(x)= {\sum_{n=0}^{\infty} a^n \cos (b^n \pi x)},
\end{equation}
where $a=0.7$ and $b=10.0$ and $x \in [0,1]$. 

Since we will be generating computed numbers we cut off the sum at a number $N$ for two cases: (1) $N=2$ to generate numerically differentiable function variables and (2) $N=20$ for numerically non-differentiable function variables. 

\begin{figure}[h]
\centering
\includegraphics[scale=0.35]{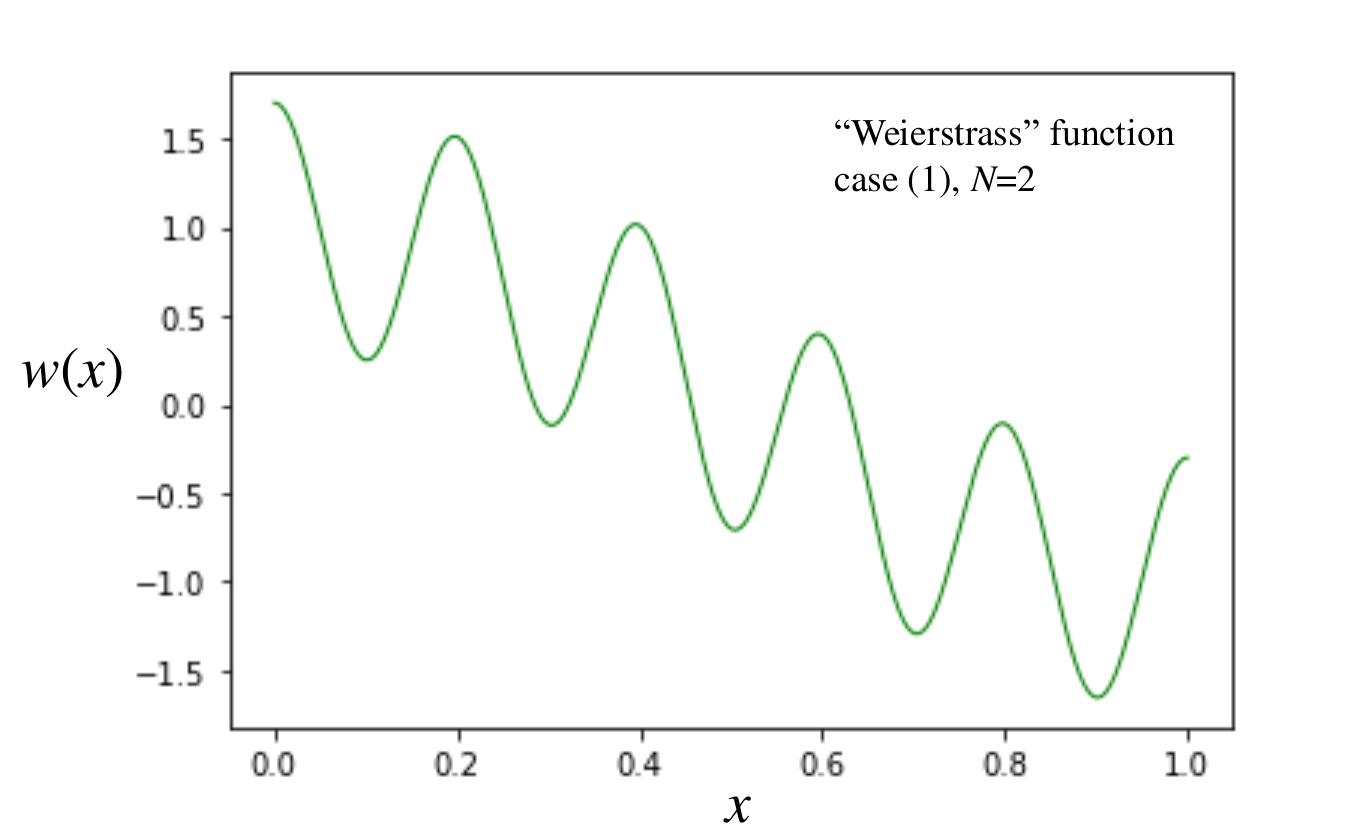}
\caption{\label{fig:Weier-smooth} Smooth version of the "Weierstrass function," $N=2$ version. }
\end{figure}

\begin{figure}[h]
\centering
\includegraphics[scale=0.32]{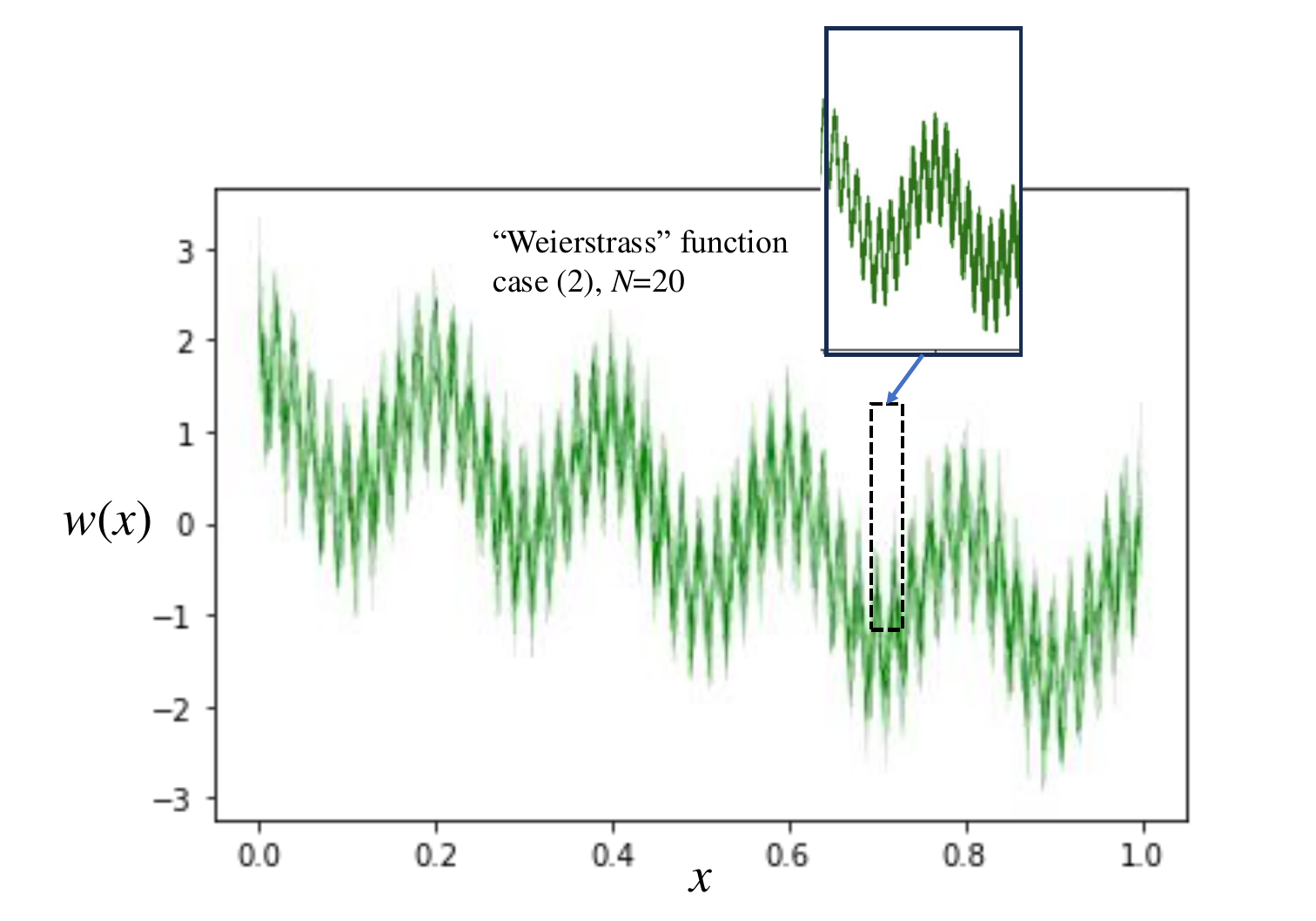}
\caption{\label{fig:Weier-fractal} Fractal version of the Weierstrass function, $N=20$ version with small portion of $w(x)$ expanded to show more structure in small regions. }
\end{figure}

The plot of case (1) "Weierstrass" function is shown in Fig. \ref{fig:Weier-smooth} and it is easy to see that it is smooth, i.e. non-fractal. 

For case (2) the large $N$ value guarantees that the fluctuations in the trigonometric function are below the computer double precision ($\sim 10^{-16}$) and the minimum spacing between points in the numerics below is roughly $10^{-4}$. In this case the numerical calculations will "see" the function values as continuous, but non-differentiable. Fig. \ref{fig:Weier-fractal} shows this in some detail.

\begin{figure}[h]
\centering
\includegraphics[width=9cm]{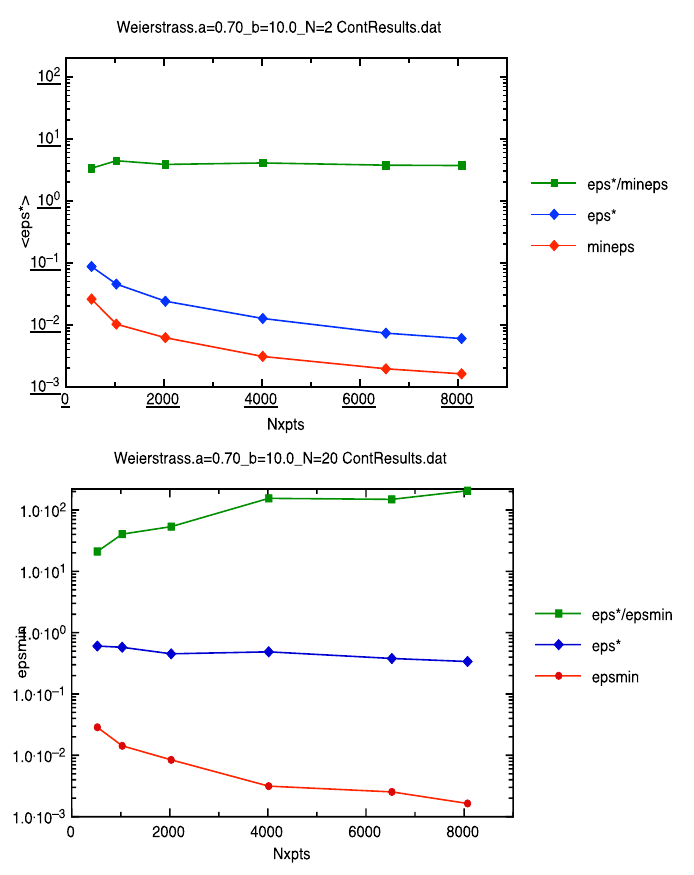}
\caption{\label{fig:Weier-Cont} Continuity statistic for "smooth" and "fractal" versions of the Weierstrass function. The continuity statistic $\epsilon^*$ decreases with increase density of points, but not as quickly as the $\epsilon_{min}$ statistic. The differing rates of decrease are related to the fractal structure of the function. }
\end{figure}

In Fig. \ref{fig:Weier-Cont} we plot the continuity statistics for the $N=2$ and $N=20$ cases.  For the N=2 case the $\epsilon^*/\epsilon_{min}$ ratio stays constant as we add more points to the time series, i.e. increasing the density of points that are sampled for the statistic. The actual values of $\epsilon^*$ and $\epsilon_{min}$ decrease with increasing number of data points.  These results for the continuity statistic provide evidence that the $N=2$ function is continuous.

In the same figure we plot the continuity statistic for the fractal ($N=20$) Weierstrass case. Although the $\epsilon^*$ statistic does not fall as fast as $\epsilon_{min}$, it does decrease with increasing density of domain points. This suggests a continuous function, but one with "fine" structure. 

Finally, the derivative statistic for the two $N$ cases show that the $N=2$ case behaves like a differentiable function, but the $N=20$ case has large errors (near 50\%), which do not decrease at all even was we add an order of magnitude more points to the same interval.

\begin{figure}[h]
\centering
\includegraphics[scale=0.5]{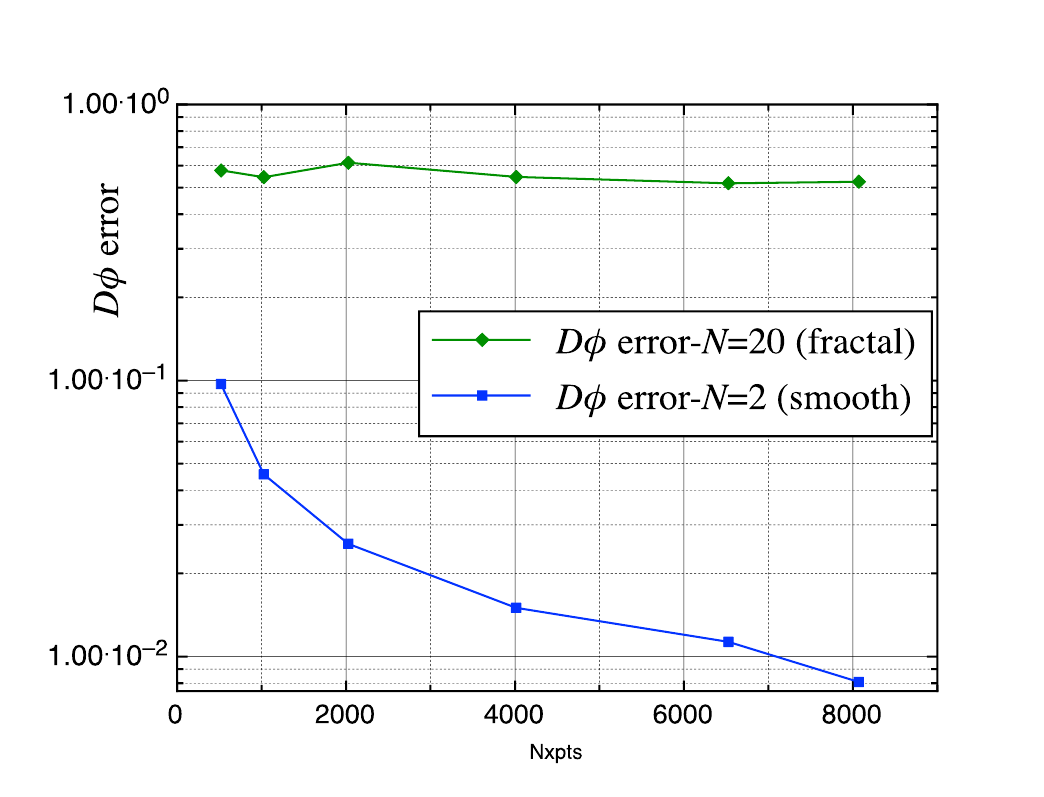}
\caption{\label{fig:Weier-Dtv} The Derivative statistic for smooth and fractal Weierstrass functions. The $N=2$ smooth Weierstrass $D\phi$ error asymptotically becomes linear and small as one would expect when the nearest neighbors of the function points are well approximated with the slope. But the $N=20$ Weierstrass $D\phi$ error starts out at an error of 57\% (a large error) and never deviates from values in that range, implying that the true Weierstrass function is not differentiable. }
\end{figure}

\end{document}